\documentclass[%
 reprint,
superscriptaddress,
 amsmath,amssymb,
 aps,
prb,
floatfix
]{revtex4-2}

\usepackage{graphicx}
\usepackage{dcolumn}
\usepackage{bm}
\usepackage{units}
\usepackage{hyperref}
\usepackage{xcolor}
\usepackage{xspace}
\usepackage{algorithm}
\usepackage{algpseudocode}

\renewcommand{\selectlanguage}[1]{}

\newcommand{\change}[1]{{\color{black}{#1}}}

\begin{document}

\title{Automatic Identification of Traps in Molecular Charge Transport Networks of Organic Semiconductors}

\author{Zhongquan Chen}
\author{Pim van der Hoorn}
\author{Bj\"orn Baumeier}
\affiliation{Department of Mathematics and Computer Science \& Institute for Complex Molecular Systems, Eindhoven University of Technology}

\date{\today}

\begin{abstract}
This paper introduces a method to identify traps in molecular charge transport networks as obtained by multiscale modeling of organic semiconductors. Depending on the materials, traps can be defect-like single molecules or clusters of several neighboring ones, and can have a significant impact on the dynamics of charge carriers. Our proposed method builds on the random walk model of charge dynamics on a directed, weighted graph, the molecular transport network. It comprises an effective heuristic to determine the number of traps or trap clusters based on the eigenvalues and eigenvectors of the random walk Laplacian matrix and uses subsequent spectral clustering techniques to identify these traps. In contrast to currently available methods, ours enables identification of trap molecules in organic semiconductors without having to explicitly simulate the charge dynamics \change{and is applicable to a variety of energy- or topology-based traps in homomolecular or mixed systems with or without detailed-balance}. As a prototypical system we \change{simulate} an amorphous morphology of bathocuproine, a material with known high energetic disorder and charge trapping. Based on a first-principle multiscale model, we first obtain a reference charge transport network and then \change{purposefully} modify its properties to represent different trap characteristics. In contrast to currently available methods, our approach successfully identifies both single trap, multiple distributed traps, and a combination of a single-molecule trap and trap regions on an equal footing. 
\end{abstract}
\maketitle

\section{Introduction}
\label{Intro}
Organic semiconductors (OSCs) are materials composed of organic molecules that are often organized in a disordered, amorphous structure, and exhibit semiconducting properties. Unlike traditional inorganic semiconductors, OSCs are flexible and allow for much easier tuning of charge mobility, so they have found applications in sensing devices~\cite{liao_organic_2013}, high-performance computing~\cite{van_de_burgt_organic_2018}, organic light-emitting diodes~\cite{PFEIFFER200389}, and organic photovoltaic cells~\cite{hains_molecular_2010}. The functionality and controllable charge mobility of OSC are to a large extent credited to so-called traps, which at the microscopic level are the molecules that can be occupied by charge carriers resulting in a significant change of charge mobility~\cite{li_universal_2014,haneef_charge_2020,olthof_ultralow_2012,witt_high_2024}. Those traps are usually single molecules, or a region consisting of very few molecules. Charge carriers can easily occupy those trapping molecules, while altering external conditions likely results in the release of carriers from the traps. Such behaviors lead to sensitive and controllable charge mobility of OSC. 

A wide range of OSC applications~\cite{kim_pattern_2017,kim_modulation_2020,van_de_burgt_non_2017,kim_emerging_2020,keene_enhancementmode_2020} revealed that by tuning the number of charge carriers one can achieve controllable charge mobility. In trap dominated materials where the carrier number is greater than that of traps, only a portion of the carriers is captured by the traps and the remaining carriers can experience fast transport. For example,~\cite{coehoorn_charge_2005,pasveer_unified_2005} show that in the Gaussian disorder models used for the theoretical study of charge transport in OSC, when the carrier number is increased by a factor two, the mobility can increase by approximately 100 times. \change{Intriguingly, traps also play an important role for the conduction mechanism in inorganic electronic materials, e.g., in the context of trap-assissted tunneling~\cite{10.1063/1.3624472,9829231}.}

Zooming into the molecular resolution, charge transport in OSCs is a sequence of transition events between the localized states~\cite{coropceanu_charge_2007, Baumeier2011} and is modeled as a continuous time random walk (CTRW) process~\cite{doyle_random_2000}. The transition rates of carriers depends on all the individual molecules' geometries and relative orientation, which affect electronic structure properties such as the  energy levels, electronic coupling elements between the molecules, and reorganization energies. Those quantities can be calculated from a first-principle multiscale model detailed in Section~\ref{sec:MSM}. 

On a macroscopic level, traps are often considered in the literature in terms of the energy density of state (DOS) $p(E)$, typically assumed to be Gaussian curves or exponential. In equilibrium, the mean energy of a charge carrier in the DOS is
\begin{equation}
    E_{\infty}=\frac{\int_{-\infty}^{\infty} E g(E) p(E) dE }{\int_{-\infty}^{\infty} g(E) p(E) dE }.
    \label{equ:einf}
\end{equation}
Here $g(E)=[\exp(\frac{E-E_F}{k_B T}) + 1]^{-1}$ is the Fermi-Dirac distribution with the Fermi energy $E_F$ determined by $\int_{-\infty}^{\infty} g(E) p(E) dE = N_c $, with $N_c$ being the number of charge carriers. Molecules with energies much lower than $E_{\infty}$ are then considered as (deep) traps. However, such a qualitative criterion is insufficient to identify traps in a molecular charge transport network for several reasons: First, the estimate of $E_{\infty}$ is based on a chosen model DOS which has some assumed continuous distribution. A realistic material, even on the scale \unit[100]{nm}, will, however, not exhibit such a continuous DOS. Second, a discrete version of Eq.~\eqref{equ:einf} depends on the number of molecules in the system, and the equilibrium energy in such a discrete DOS is dependent on system size~\cite{PhysRevB.82.193202}. Third, focusing on the DOS alone ignores other contributing factors to the charge dynamics, or the features of the transport network, such as electronic coupling elements between pairs of neighboring molecules, structural details of the material and or spatial correlations. These details are connected to the variety of physical sources for traps, e.g., interfacial effects, defects in molecular packing, or chemical impurities. \change{In some of these cases, going from the macroscopic DOS to the molecule-specific information allows considering molecules with high Boltzmann occupation probabilities in equilibrium $p_i = \exp(\beta E_i)/\sum_i\exp(\beta E_i)$ (with $\beta = (k_\text{B}T)^{-1}$, $E_i$ the energy of molecule $i$, and $T$ the temperature) as traps. However, besides being similarly problematic as the DOS in defining a threshold value to use, this approach also relies on the pairwise intermolecular transfer rates obeying detailed balance, does not cover topology-based traps, and alone is insufficient to identify trap regions in materials with strong spatial correlations.} This makes it difficult to provide a quantitative definition of traps that can be used for identification \change{of all trap characteristics on an equal footing}. 

At present, there are no methods for the identification of traps in molecular charge transport networks that perform reliably for all different trap types. Few attempts have been reported in identifying trap regions, or clusters, based on analyzing the actual simulated dynamics, e.g., via kinetic Monte Carlo (KMC)~\cite{schulze_efficient_2008}. Qualitatively, once entered into such a trap region, the random walk (representing the charge dynamic of a single carrier) transitions mostly within it and escaping it is a rare event, making such KMC simulations very time-consuming. Two methods to accelerate KMC simulations which indirectly involve trap identification have previously been discussed. One is based on the (stochastic) watershed algorithm filling regions (''basins'') in the spatially resolved energy distribution~\cite{brereton_efficient_2014}. This purely energy-based criterion does, however, not consider additional details of the factors influencing the molecular charge transport network. The second method~\cite{stenzel_general_2014} is based on a graph-theoretic decomposition (GD) and makes use of the fact that in the presence of trapping regions the Markov chain on the molecular charge transport network is nearly completely decomposable~\cite{stewart_introduction_1994}, allowing the associated graph to be partitioned into subgraphs. While this method takes the full information of the hopping-type dynamics into account, it is sensitive to the choice of parameters (related to, e.g., graph connectivity properties or transition rate ratios) and is not successful in identifying single trap nodes in the graph (as we will also discuss in Section~\ref{sec:GD}). 

In this paper, we propose a new method that builds upon the idea of graph partitioning by using spectral clustering based on a specific type of Laplacian matrix of the graph~\cite{hagen_new_1992,von_luxburg_tutorial_2007}. The aim of this method is to separate the graph into partitions, by minimizing a normalized cut cost function, such that the random walk processes rarely transitions between different partitions. While obtaining a minimized normalized cut is a NP-hard problem, a relaxed solution of this discrete optimization problem can be obtained from the eigenvectors of the random walk Laplacian matrix which will be introduced in Section~\ref{subsec:SCT}. Our proposed method includes an effective heuristic to determine the number of traps or trap clusters based on these eigenvalues and eigenvectors of this random walk Laplacian and subsequent performing spectral clustering (using K-means clustering) to identify the traps. The former depends on a single threshold parameter for which we find an optimal choice \change{nearly} independent of the specific system.

Using the charge transport network resulting from a multiscale model of an amorphous morphology of bathocuproine (BCP)~\cite{tsung_carrier_2008}, a molecular material with known high energetic disorder~\cite{kaiser_novo_2021}\change{, low charge mobility~\cite{10.1063/5.0049513},} and complex charge trapping behavior, \change{as a baseline with subsequent modifications} we demonstrate that our approach successfully identifies \change{energy-based} single traps, multiple distributed traps, and a combination of a single-molecule trap and trap regions on an equal footing. \change{It is furthermore shown that the approach is robust under application of external electric fields and also detects topology-based traps and traps in mixed-molecular materials that are not energy based without modifications to the algorithm.} We also find a strong relation between the cost function associated with the normalized cut and the charge-carrier dynamics simulated in a time-of-flight setup~\cite{lebedev_charge_1997,chen_graph_2024}, as well as the physical characteristics of the trap (regions). 

In what follows, Section~\ref{sec:MSM} will introduce the elements of the first-principle multiscale model used to obtain the molecular charge transport network of BCP based on a combination of classical molecular dynamics (MD) with quantum electronic structure theory on the level of density-functional theory (DFT), and the calculation of the time-of-flight to assess charge-carrier dynamics of the model. In Section~\ref{subsec:SCT} we give the details of the spectral-clustering based trap identification method we propose in this work, including the determination of the cluster number and K-means clustering. The results of the application of this method to the BCP system and its modifications to cover different trap types is presented and discussed in Section~\ref{sec:result}. A brief conclusion and discussion concludes the paper. 

\section{Multiscale Model}
\label{sec:MSM}
To create the \change{baseline} molecular charge transport network of BCP, we employ a multiscale model which connects morphology simulations based on classical MD with quantum-classical methods to evaluate the Marcus rate for electron transfer between two molecules $i$ and $j$:
\begin{equation}
    \omega_{ij} = \frac{2\pi}{\hbar} \frac{|J_{ij}|^2}{\sqrt{4\pi R_{ij} k_\text{B}T}} \exp\left(-\frac{(E_{ij} + q \vec{F} \cdot \vec{r}_{ij} - R_{ij})^2}{4R_{ij} k_\text{B}T}\right) ,
    \label{equ:Marcus}
\end{equation}
where $\hbar$ is the reduced Planck constant and $k_\text{B}$ the Boltzmann constant. The temperature $T$ (in \unit[]{K}), the charge of the carrier $q$ (in \unit[]{e}) and the external electric field $\vec{F}$ (in V/m) can be considered the external, environmental parameters of the simulation. The vector $\vec{r}_{ij} = (r^x_{ij},r^y_{ij},r^z_{ij})^\text{T}$ connects the center-of-masses (COMs) of molecules $i$ and $j$, which is our setup  calculated using cyclic boundary conditions in the Cartesian $x$- and $y$-directions. The remaining physical, material-specific (or rather transfer-pair-specific) quantities are the reorganization energy $R_{ij}$, the electronic coupling $J_{ij}$, and the energy difference $E_{ij} = E_i - E_j$ (all in \unit[]{eV}). With the information about the molecular COMs and the rates, one can construct the graph $\mathbf{G}=(\mathbf{V}, \mathbf{W})$ on which the CTRW takes place. Here each node $i \in \mathbf{V}$ corresponds to a molecule and $\mathbf{W}: \omega_{ij}$ is the weighted adjacency matrix represented by the Marcus rates. A rate is only considered between molecules $i$ and $j$ if their closest-contact distance is smaller than $r_\text{cutoff}=\unit[0.5]{nm}$. 

\subsection{Molecular Dynamics}
Atomistic molecular dynamics simulations of bulk amorphous BCP are conducted using the GROMACS software package~\cite{berendsen_gromacs_1995}, employing a gromos54a7 type force field obtained via the tool Automated Topology Builder~\cite{stroet_automated_2018}. Initially, 1000 BCP molecules are randomly placed in a cubic cell with a side length of \unit[10]{nm}. Periodic boundary conditions are applied throughout in all three spatial directions. After energy minimization, the system is simulated for \unit[1]{ns} at a constant temperature of \unit[300]{K} and constant pressure of \unit[1]{bar} in the $NpT$ ensemble using the V-rescale thermostat~\cite{bussi_canonical_2007} with the coupling time constant \unit[0.1]{ps} and the Parrinello-Rahman barostat~\cite{parrinello_polymorphic_1981} with a time constant for pressure coupling \unit[2]{ps}. The leap-frog algorithm~\cite{van_gunsteren_leap-frog_1988} is used to integrate the equation of motion with a time step of \unit[1]{fs}. 

In the next step, we employ simulated annealing to first increase the temperature to \unit[800]{K} during a period of \unit[0.5]{ns}, i.e., well above the glass transition temperature of the material. The system is maintained at the temperature for \unit[1]{ns} before cooling back down to \unit[300]{K} during a period of \unit[0.5]{ns}. Such heating-cooling cycle is repeated three times. After this simulated annealing, a production run is conducted for \unit[2]{ns} using the $NpT$ ensemble. The final configuration of BCP is chosen for the further steps in the multiscale model, whose configuration is a cubic box with a length of \unit[8.1]{nm} and a density of $\unit[1.13]{g/cm^3}$. This is consistent with the experimentally measured value of $\unit[1.12]{g/cm^3}$~\cite{xiang_method_2007} and another reported MD result of $\unit[1.16]{g/cm^3}$~\cite{degitz_simulating_2022}. 

\subsection{Electronic Structure Calculations} 
\label{sec:es}

Effective single-electronic wave functions $\phi_l (\vec{r})$ and associated energies $\epsilon_l$ for a system with $N_\text{el}$ electrons are determined as solutions to the Kohn--Sham equations~\cite{kohn_self_1965}
\begin{equation}
    \begin{split}
    \left(-\frac{1}{2}\nabla^2_{\vec{r}} + v_\text{ext}(\vec{r}) + v_\text{H}[\rho](\vec{r}) + v_\text{XC}[\rho](\vec{r})\right) \phi_l(\vec{r}) \\= H^\text{KS} \phi_l(\vec{r}) = \epsilon_l \phi_l (\vec{r}) ,
    \end{split}
    \label{eq:KS2}
\end{equation}
where $v_\text{ext}$ in an external potential (typically from the nuclei), $v_\text{H}[\rho]$ the electrostatic Hartree potential of a classical charge density $\rho(\vec{r})$, and $v_\text{XC}[\rho]$ the exchange-correlation functional containing explicit quantum-mechanical electron-electron interactions. The charge density is determined from the single-particle wave functions as $\rho(\vec{r})=\sum\limits_{l=1}^{N_\text{el}} \left\vert\phi_l(\vec{r})\right\vert^2$. As the Hartree and exchange-correlation potential depend on the thus defined density, solutions to Eq.~\ref{eq:KS2} have to be found self-consistently. This corresponds to finding the ground-state density $\rho_0$ that minimized the total energy of the system
\begin{equation}
    U[\rho] = T_s[\rho] + \int v_\text{ext}(\vec{r}) \rho(\vec{r}) d \vec{r} + E_\text{H}[\rho] + E_\text{XC}[\rho]
    \label{eq:KS_model}
\end{equation}
where $T_s[\rho]$ is the kinetic energy, $E_\text{H}[\rho]$ and $E_\text{XC}[\rho]$ the Hartree and exchange-correlation energies, respectively. 

The practical calculations in this work have been performed with the ORCA software~\cite{Neese2012a} using the BHANDHLYP exchange-correlation functional~\cite{rudberg_kohnsham_2011} with the def2-tzvp~\cite{weigend_accurate_2006} basis set to represent the Kohn--Sham wave functions $\phi_l(\vec{r})$.

\subsection{Calculation of Marcus Rates}
Equations~\ref{eq:KS2} and~\ref{eq:KS_model} can be solved for different total charge states $x=\text{n,c}$ and corresponding  equilibrium geometries $X=\text{N,C}$, where $\text{n}$ and $\text{N}$ stand for ''neutral'' and $\text{c}$ and $\text{C}$ for ''charged''. The respective total energies will be denoted in the following as $U^\text{xX}$, dropping the explicit mention of the functional dependency on $\rho$ for compactness. The reorganization energy $R_{ij}$ for charge transfer from molecule $i$ to molecule $j$ in the Marcus rate Eq.~\ref{equ:Marcus} \change{is a sum of contributions from structural changes after molecule $i$ donates the charge ($R_i^\text{n}$) and molecule $j$ accepting the charge ($R_j^\text{c}$)~\cite{Baumeier2011,10.1063/5.0049513}}. It is calculated as
\begin{equation}
    R_{ij} = R_{i}^\text{n} + R_j^\text{c} = U_i^\text{nC} - U_i^\text{nN} + U_j^\text{cN} - U_j^\text{cC}.
\end{equation}
Specifically for BCP, we obtain the value $R_{ij}=\unit[0.49]{eV}$\change{, which is considered a single-molecule quantity in homo-molecular materials}.

\change{For the remaining quantities that depend on the material morphology (site energies and electronic coupling), special care is taken to avoid discrepancies between the molecular structures obtained via classical MD and the DFT calculations. To remove bond length fluctuations introduced by molecular dynamics simulations, as they are already integrated out in the derivation of the rate expression, molecular fragments with rigid, planar $\pi$ systems are substituted by DFT-optimized versions~\cite{Baumeier2011}.} The site energy $E_i = E_i^\text{c} - E_i^\text{n}$ is the difference between the total energies of the system in which molecule $i$ is carrying a charge or not, corresponding to the ionization potential in case of hole transport and the negative of the electron affinity in case of electron transport. These total energies in turn consist of different contributions associated with different physical mechanism, i.e.,
\begin{equation}
E_i^x = U_i^{xX} + E_i^{x,\text{static}} + E_i^{x,\text{polar}},
\label{eq:Es}
\end{equation}
where $U_i^{xX}$ is the intramolecular contribution from single molecules as above, while $E_i^{x,\text{static}}$ and $E_i^{x,\text{polar}}$ are contributions arising from intermolecular interactions.  

In the multiscale approach used in this work, these contributions are calculated with a microelectrostatic model using parametrized point charge representations for the molecular electrostatic potential~\cite{https://doi.org/10.1002/jcc.540110311} and atomic dipole polarizabilities to model the self-consistent response of the molecules to the associated fields~\cite{thole_molecular_1981}. Long-range electrostatic interactions are accounted for via a periodic embedding of aperiodic excitations based on Ewald summation~\cite{poelking_impact_2015, poelking_long-range_2016}. Polarization effects are considered within a cutoff of \unit[4.0]{nm} around each individual molecule. Practical calculations of the site energies are performed using the VOTCA software~\cite{Baumeier2011,doi:10.1021/acs.jctc.8b00617,10.1063/1.5144277,Baumeier2024}. The resulting site energy distribution for BCP is shown in Figure~\ref{fig:E_J}(a) in~\ref{app:es}.

The coupling element $J_{ij}$ between molecule $i$ and $j$ is calculated using the Dimer-Projection Method~\cite{baumeier_density_2010}. In the case of hole transport, it uses the Kohn--Sham wave functions of the highest-occupied molecular orbital (HOMO) of the isolated molecules (monomers) $\phi^\text{HOMO}_i(\vec{r})$ and $\phi^\text{HOMO}_j(\vec{r})$, the Kohn--Sham Hamiltonian $H^\text{KS}_\text{D}$ (see Eq.~\ref{eq:KS2}) of the dimer and the associated full set of wave functions $\left\{ \phi_\text{D}(\vec{r})\right\}$. Specifically, the coupling element is determined as: 
\begin{equation}
    J_{ij} = \frac{ J^0_{ij}- \frac{1}{2}(e_i+e_j) S_{ij} }{ 1- S_{ij}^2 }
    \label{equ:JAB}
\end{equation}
where $J^0_{ij} = \langle \phi_i | \hat{H}^\text{KS}_\text{D} | \phi_j \rangle $, $e_i = \langle \phi_i | \hat{H}^\text{KS}_\text{D} | \phi_i \rangle $, $e_j = \langle \phi_j | \hat{H}^\text{KS}_\text{D} | \phi_j \rangle $, and $S_{ij}=\langle \phi_i | \phi_j \rangle $ with bra-ket notation. The Hamiltonian of the dimer,  $H^\text{KS}_\text{D}$ (see Eq.~\ref{eq:KS2}), is diagonal in its eigenbasis $\left\{\vert \phi^\text{D}_k\rangle\right\}$ with eigenvalues $\left\{ \epsilon^\text{D}_k\right\}$, so $H^\text{KS}_\text{D} = \text{diag}(\epsilon^\text{D})$. With the projections of the monomer functions on the dimer eigenbasis, i.e., $p_{ik} = \langle \phi_i | \phi^\text{D}_k \rangle$ and  $p_{jk} = \langle \phi_j | \phi^\text{D}_k \rangle$, $J^0_{ij}$ can be calculated as $J^0_{ij} = \mathbf{p}_i^\text{T} \text{diag}(\epsilon^\text{D}) \mathbf{p}_j$. Similarly, $e_{i(j)} = \mathbf{p}_{i(j)}^\text{T} \mathbf{p}_{i(j)}$ and $S_{ij} =  \mathbf{p}_i^\text{T} \mathbf{p}_j$. All  of these operations are performed in the basis set representation of the Kohn--Sham wave functions (see Sec.~\ref{sec:es}) as implemented in VOTCA. The resulting distribution of coupling elements for BCP is shown in Figure~\ref{fig:E_J}(b) in~\ref{app:es}.

\subsection{Time-of-flight Calculation}
\label{sec:tof}
After computation of the Marcus rate from the multiscale model and defining the graph $\mathbf{G}=(\mathbf{V},\mathbf{W})$, the charge dynamics can be modeled as a continuous-time random walk process on the graph. To model the Source-Sink conditions, some vertices are used as \emph{Source} to represent the electrode where charge carriers are injected, and some vertices as \emph{Sink} where charge carriers are detected and time-of-flight (ToF) is recorded. \change{Specifically, we consider as source vertices the molecules with a COM component $r_i^x < \unit[0.5]{nm}$, and as sink vertices those with $r_i^x > \max(r_i^x)-\unit[0.5]{nm}$. As a result, the graph retains the cyclic boundary conditions only in the Cartesian $y$- and $z$-directions.}

\change{In the model,} each node can be occupied by at most one charge carrier, so a system with $N$ molecules and $N_c \le N$ charge carriers has a total of ${N \choose N_c}$ occupation situations. \change{While spin is not explicitly taken into account, the assumption of such a Pauli-like exclusion is a computationally effective way to avoid the calculation of explicit long-range Coulomb repulsion between electrons at different sites. Note that the formation of double excitations or bi-polarons is not included 
in the main transport dynamics model and requires different rate formulations.} Each of these occupation situations is considered as a state $\mathbf{s}$. If all the carriers' occupations are in a Source node, the state is called a Source state, and if at least one of the Sink nodes is occupied, the state is called a Sink state. The transition rates between the states can be obtained from the matrix $\mathbf{W}:\omega_{ij}$, since the connectivity of states is encoded in the connectivity of the nodes (molecules), as detailed in ~\cite{chen_graph_2024}. In particular, the transition rate $\Omega_{\mathbf{s} \mathbf{s}' }$ from state $\mathbf{s}$ to $\mathbf{s}'$ is:
\begin{equation}
\label{eq:transition_rates}
	\Omega_{\mathbf{s} \mathbf{s}^\prime} =
	\begin{cases}
	     0			&  \mathbf{s} \text{ is not connected to } \mathbf{s}^\prime,\\
	    \omega_{ij}	&  \mathbf{s} \text{ is connected to } \mathbf{s}^\prime \text{ due to } (i,j).
	\end{cases}
\end{equation}

Then the transition probability from state $\mathbf{s}$ to $\mathbf{s}'$ is $p_{\mathbf{s} \mathbf{s}^\prime} = \Omega_{\mathbf{s} \mathbf{s}^\prime}/D_\mathbf{s}$ where $D_\mathbf{s} := \sum_{\mathbf{s}^\prime \ne \mathbf{s}} \Omega_{\mathbf{s} \mathbf{s}^\prime}$.
And the expected time from state $\mathbf{s}$ to reach the Sink state $\tau_\mathbf{s}$ is calculated 
~\cite{chen_graph_2024} via: 
\begin{equation}
	\tau_\mathbf{s} = \begin{cases}
		\frac{1}{D_\mathbf{s}} + \sum_{\mathbf{s}^\prime \ne \mathbf{s}} p_{\mathbf{s} \mathbf{s}^\prime} \tau_{\mathbf{s}^\prime} &\text{if $\mathbf{s}$ is not a sink state},\\
		0 &\text{else.} 
	\end{cases}
    \label{eq:hitting_time}
\end{equation} 

Since we need to consider all possible starting nodes of the carriers, all the Source states need to be considered. 
The random walk process has been modeled as a parallel electric network of capacitors~\cite{doyle_random_2000}, and accordingly, we take the ToF to be the harmonic mean:
\begin{equation}
\tau = N_\text{source} \left[\sum_{\mathbf{s}\in \text{Source}} (\tau_\mathbf{s}^\ast)^{-1}\right]^{-1},
\label{eq:tof2}
\end{equation}
with $N_\text{source}$ the number of source states. \change{Note that the above model does not include a constant supply of charge carriers at the source but instead corresponds to a simulation with a short, pulsed creation of charge carriers, equivalent to KMC or Master Equation based models~\cite{chen_graph_2024}.}

\section{Spectral Clustering Method}
\label{subsec:SCT}

In this section we present the theoretical background of spectral clustering for trap identification, including a recapitulation of details about the random walk Laplacian matrix, the graph partitioning based on its eigenvalues and eigenvectors, and the K-means clustering algorithm. We will also propose our heuristics for determining the number of traps, within this framework.

\subsection{Random Walk Laplacian Matrix and Graph Partitioning}
Identifying traps within a multiscale modeled molecular charge transport network corresponds to finding the regions of nodes in the corresponding graph in which the random walk process spends a long time, while the transition between the regions is rare. 

In spectral clustering theory, see for example \cite{von_luxburg_tutorial_2007} Proposition 5, finding such regions of nodes corresponds to cutting through the graph such that the resulting partitions $\mathbf{G}_1, \cdots, \mathbf{G}_k$ minimize the objective function
\begin{equation}\label{eq:Ncut}
    \text{NCut}(\mathbf{G}_1,\cdots,\mathbf{G}_k) := \sum\limits_{i=1}^k \frac{\text{cut}(\mathbf{G}_i,\bar{\mathbf{G}}_i)}{ \text{vol}(\mathbf{G}_i) } \mbox{,}
\end{equation}
where $\text{cut}(\mathbf{G}_i,\bar{\mathbf{G}}_i) = \frac{1}{2} (W_{\mathbf{G}_i,\bar{\mathbf{G}}_i} + W_{\bar{\mathbf{G}}_i, \mathbf{G}_i }) $, with $W_{\mathbf{A},\mathbf{B}} := \sum\limits_{i \in \mathbf{A},j \in \mathbf{B}} \omega_{ij}$, and $\text{vol}(\mathbf{G}_i)$ the volume $\mathbf{G}_i$, calculated by summing up the weights of the edges within $\mathbf{G}_i$. 

In the context of trap identification, the partition that minimizes Eq.~\ref{eq:Ncut} will correspond to the traps in the system and the remaining molecules. That is, we have $K = k-1$ traps and one element $\mathbf{G}_i$ will represent all non-trap nodes. 

The problem is that solving Eq.~\ref{eq:Ncut} is known to be a NP hard problem. However, a relaxation of it can be solved using the so-called \emph{random walk Laplacian} matrix. For a weighted graph $\mathbf{G}=(\mathbf{V}, \mathbf{W})$ define the out-degree matrix by $\mathbf{D}:=D_{i,i} = \sum_j \omega_{ij}$, and the Laplacian matrix by $ \mathbf{L} = \mathbf{D} - \mathbf{W} $. The random walk Laplacian matrix of a graph $\mathbf{G}$ is given by
\begin{equation}
    \mathbf{L}_\text{rw} = \mathbf{I} - \mathbf{D}^{-1} \mathbf{W}\mbox{,}
\label{equ:L}
\end{equation} 
So $\mathbf{L}_\text{rw}$ is the Laplacian matrix normalized by $\mathbf{D}$.

Note that the charge transport graph is directed and hence the Laplacian matrix is not symmetric in general. Nevertheless, the random walk Laplacian $\mathbf{L}_\text{rw}$ has a real-valued spectrum. This follows from the fact that 
\begin{equation}
    \mathbf{L}_\text{rw} = \mathbf{I} - \mathbf{D}^{-\frac{1}{2}} (\mathbf{I} - \mathbf{L}_\text{sym}) \mathbf{D}^{-\frac{1}{2}}\mbox{,}
\end{equation} 
where $\mathbf{L}_\text{sym} = \mathbf{I} - \mathbf{D}^{-\frac{1}{2}} \mathbf{W} \mathbf{D}^{-\frac{1}{2}}$ is the normalized symmetric Laplacian. So $\mathbf{L}_\text{rw}$ is similar to $\mathbf{L}_\text{sym}$, and since the latter has a real-valued spectrum, $\mathbf{L}_\text{rw}$ has $N$ real eigenvalues: $\lambda_1=0 \leq \lambda_2 \leq \lambda_3 \leq \cdots \leq \lambda_N$. The fact that $\mathbf{L}_\text{rw}$ has real eigenvalues enables the use of spectral clustering methods for our directed charge transport network. 

The idea behind spectral clustering is to consider the relaxation of the NP hard discrete minimization problem for NCut~\ref{eq:Ncut}, which tries to find $\mathbf{T} \in \mathbb{R}^{N \times k}$ that minimizes
\begin{equation}
    \min_{\mathbf{T} \in \mathbb{R}^{N \times k}} \text{Tr}(\mathbf{T}^{\prime} \mathbf{D}^{-\frac{1}{2}} \mathbf{L}_\text{rw} \mathbf{D}^{-\frac{1}{2}} \mathbf{T})
    \label{eq:min2}
\end{equation}
subject to the constraint that $\mathbf{T}^\prime \mathbf{T} = \mathbf{I}$. The solution $\mathbf{T}^\ast$ to Eq.~\ref{eq:min2} is formed by the first $k$ eigenvectors of the random walk Laplacian matrix $\mathbf{L}_\text{rw}$.

It should be noted that solving~\ref{eq:min2} does not yield a partition of the graph, as this is a relaxation of the true discrete optimization problem we wish to solve. To construct a partition in practice a K-means algorithm, introduced in the next subsection, is performed on the rows of the solution $\mathbf{T}^\ast$ to~\ref{eq:min2}.

\subsection{K-means Clustering Algorithm}
In general, the K-means clustering method partitions a dataset consisting of $N$ data points into distinct clusters, minimizing the distance between points in each cluster. Let $k \ge 2$ and consider $N$ data points: $\{ \mathbf{x}_1, \mathbf{x}_2, \cdots, \mathbf{x}_N \}$ where $\mathbf{x}_i \in \mathbb{R}^d$. For any partition $\mathcal{C}_1,\cdots,\mathcal{C}_k$ of $\{1,2, \dots, N\}$ we define the cost function
\begin{equation}\label{eq:cost}
Z(\mathcal{C}_1,\cdots,\mathcal{C}_k) = \sum\limits_{l=1}^{k} \frac{1}{2|\mathcal{C}_l|} \sum\limits_{i,j \in \mathcal{C}_l} || \mathbf{x}_{i} - \mathbf{x}_{j} ||_2^2\mbox{.}
\end{equation}
The objective of K-means clustering is to find the partition $\mathcal{C}^\ast_1,\cdots,\mathcal{C}^\ast_k$ that minimizes Eq.~\ref{eq:cost}. That is, the partition minimizes the pairwise squared distance within each cluster normalized by the cluster size. A practical approach to this is via a local search algorithm such as Lloyd's algorithm~\cite{lloyd_least_1982}:
\begin{algorithm}[H]
    \caption{Lloyd's algorithm for K-means clustering}\label{alg2}
    \begin{algorithmic}[1]
        \State \texttt{Input $k$ and $\mathbf{x}_1,\mathbf{x}_2,\cdots,\mathbf{x}_N$, $\mathcal{C}_1,\cdots,\mathcal{C}_k$ are randomly initialized}
        \While {\texttt{not converged}}
        	\State \texttt{Compute $c_l = \frac{1}{|\mathcal{C}_l|} \sum\limits_{i \in \mathcal{C}_l} \mathbf{x}_{i}$ for $l = 1,2,\cdots,k$}
        	\State \texttt{Update $\mathcal{C}_1,\cdots,\mathcal{C}_k$ by assigning each point $\mathbf{x}_{i}$ to the cluster whose  centroid $c_l$ is closest to.}
        	\State \texttt{If the cluster assignment did not change, convergence is achieved.} 
        \EndWhile
        \State \texttt{Return cluster assignment $\mathcal{C}_1,\cdots,\mathcal{C}_k$.}
    \end{algorithmic}
\end{algorithm}

In this work, the input $N$ data points are the $N$ rows of the matrix $\mathbf{T}^\ast$ that is the solution to Eq.~\ref{eq:min2}. The resulting $k$ clusters are then understood to correspond to $K = k-1$ traps in the system and the remaining group of nodes representing the rest of the system.


\subsection{Determination of the Cluster Number}
\label{sec:3.1}
\begin{figure*}[tb]
    \centering
    \includegraphics[width=0.98\textwidth]{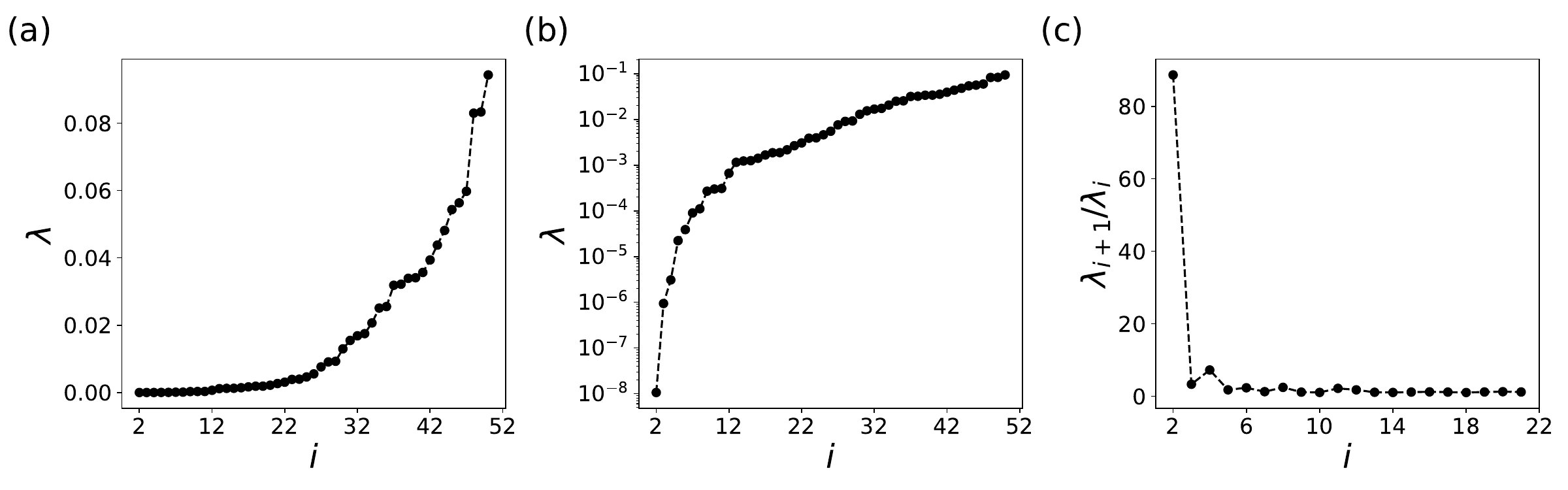}
    \caption{The first fifty eigenvalues of the random walk Laplacian matrix 
    $\mathbf{L}_\text{rw}$ (a) without log scale and (b) with log scale. The first eigenvalue is zero and is not shown in the plot. (c) Eigenvalue ratio $\lambda_{i+1}/\lambda_i$ as a function of the index.}
    \label{fig:50eigval}
\end{figure*}

For the description of the K-means clustering algorithm it follows that one needs to provide an input $k$ for the number of clusters. The main problem, as is also the case for our intended application, is that it often unclear what $k$ should be. Therefore, in this subsection will introduce an algorithm to determine the number of clusters, which is based on the more commonly applied eigengap heuristic.

Starting from spectral clustering theory, we know that the multiplicity of the smallest eigenvalue 0 of a graph Laplacian equals the number of connected components in the graph $\mathbf{G}$. If $\mathbf{G}_1, \cdots, \mathbf{G}_k$ are the connected components, then the eigenspace corresponding to eigenvalue 0 is spanned by the vectors $\{ \mathbf{1}_{\mathbf{G}_i} \}_{i=1}^{k}$. Here $\mathbf{1}_{\mathbf{G}_i}$ is the vector where elements corresponding to the nodes in cluster $\mathbf{G}_i$ are one and the rest are 0. This shows that both the eigenvalues as well as the corresponding eigenvectors contain relevant information about the possible number of clusters.

Of course, in many applications, the graph is a single connected graph and the interesting clusters do not reside in disconnected components. Hence, we cannot simply use the multiplicity of the first eigenvector.  Instead, a variety of ways to determine the number of clusters have been discussed previously. For instance, Fraley and Raftery proposed finding the number of clusters based on the log-likelihood of the data~\cite{fraley_model-based_2002}, while Still and Bialek suggested information-theoretic criteria based on the ratio of within cluster and between cluster similarity~\cite{still_how_2004}. Tibshirani {\it el at.} used gap statistics on general data points to find the cluster number~\cite{tibshirani_estimating_2001}. Specifically, in the heuristic eigengap method based on perturbation theory $k$ is chosen such that $\lambda_1,\cdots,\lambda_k$ are very small and $\lambda_{k+1}$ is relatively large~\cite{von_luxburg_tutorial_2007}. In practice, it can however be difficult to implement this heuristic as it is not well-defined what ``relatively large`` is. This problem can clearly be seen in the eigenvalues of $\mathbf{L}_\text{rw}$ for the BCP model system, as shown on linear and log scale in Figure~\ref{fig:50eigval}(a) and Figure~\ref{fig:50eigval} (b), respectively. 

To overcome the non-obvious challenge of finding a good $k$, we notice that the ratio of the eigenvalue $\lambda_{i+1}/{\lambda_i}$ is large for small indices $i$, as shown in Fig.~\ref{fig:50eigval}(c). As the index $i$ becomes large, ${\lambda_{i+1}}/{\lambda_i}$ becomes small. This prompts us to utilize the index $\ell$ yielding the maximum ${\lambda_{\ell+1}}/{\lambda_\ell}$ as a good first guess for the number of clusters.

Additionally, the eigenvectors also contain important information about possible clusters. For example, when considering 2 clusters, the entries below 0 in the second eigenvectors will correspond to nodes in one cluster while those above 0 correspond to nodes in the other cluster. From this, the idea is that we should only consider those nodes $i$ such that the $i$-th entry in the $\ell$-th eigenvector is large in absolute value. 

All together, our method first chooses the index $\ell$ such that ${\lambda_{\ell+1}}/{\lambda_{\ell}}$ is the largest. Then using the eigenvectors corresponding to the first $\ell$ eigenvalues, we consider the elements sufficiently distinct from zero. This is controlled via a parameter $a$. For each of the first $\ell$ eigenvectors, we will look at the induced subgraph on those nodes whose entries are larger than $a$ in absolute value. We then collect the disconnected components of the corresponding induced subgraph as separate graphs. After this we have a collection of small subgraphs from which we will remove all subgraphs that are themselves a subgraph of another graph in this collection. This procedure leaves us with a list of disjoint subgraphs given by the first $\ell$ eigenvectors and we take $k$ to be the number of these graphs plus 1. The full procedure to determine the cluster number $k$ is summarized in Algorithm~\ref{alg1}.

\begin{algorithm}[H]
    \caption{Determination of cluster number $k$}\label{alg1}
    \begin{algorithmic}[1]
    \State Input: $\mathbf{G}$, $\mathbf{L}_\text{rw}$ $a$.
    \State \texttt{Calculate $(\lambda_i, \mathbf{u}_i)$ for $i=1,\cdots,N$, such that $\lambda_1 \leq \lambda_2 \leq \cdots \leq \lambda_N$. }
    \For{\texttt{ $i=2,3,\cdots,N-1$ }}
    \State $\mathbf{u}_i \gets |\mathbf{u}_i|/\text{max}(|\mathbf{u}_i|)$
    \EndFor
    \State \texttt{Calculate $\ell=\text{argmax}(\frac{\lambda_{i+1}}{\lambda_i})$}

    \State Denote empty set $B=\emptyset$
    \For{\texttt{ $i=1,2,\cdots,\ell$ }}
        \State \texttt{Set node list $Q = \emptyset$}
        \For{ $j = 1,2,\cdots,N$ }
            \If {$\mathbf{u}_{ij} > a $}
                \State \texttt{add $j$ to $Q$}
                \EndIf
            \EndFor
        \State \texttt{Let $H_1, \dots, H_M$ denote the disconnected components of the induced subgraph in $G$ on the nodes in $Q$.}
        \State \texttt{Update $B = B \cup \{H_1, \dots, H_M\}$}
    \EndFor

    \For{\texttt{$H, H^\prime \in B$}}
        \If {$H \subseteq H^\prime$}
            \State \texttt{remove $H$ from $B$}
        \EndIf
    \EndFor

    \State \texttt{$k = |B|+1$}
    \end{algorithmic}
\end{algorithm}

Let us make a few remarks about this algorithm. Firstly, to select $\ell$ such that ${\lambda_{\ell+1}}/{\lambda_\ell}$ is maximum, the algorithm starts from $i=2$ since the first eigenvalue is always 0. This is also the only eigenvalue equal to 0 as we only work with connected graphs. We then normalize the eigenvectors corresponding to the selected eigenvalues as ${|\mathbf{u}_i|}/{\text{max}(|\mathbf{u}_i|) }$. The purpose for this normalization is to make the algorithm more robust and less dependent on the parameter $a$, which is used to indicate the entries of eigenvector elements distinct from zero. In particular, we can take $0 < a \leq 1$. In general, the value of $a$ should not be too close to 1 to avoid missing relevant subgraphs. It should also not be too close to zero because then it will not be selective enough. In this work we pick different increasing values of $a$ starting at $0.9$, to check how robust the analysis is for $a$ closer to $1$. Finally, note that when counting the total number of induced subgraphs, if the subgraphs are contained in other subgraphs, those do not contribute to the value of $k$.

\section{Results}
\label{sec:result}
We begin with a short semi-quantitive analysis of the effect of traps in a molecular charge transport network. As the name suggests, once a charge carrier encounters a trap, it will spend a significant amount of time in it and the observed charge dynamics will be slow (large ToF). However, if there are more carriers in the system than traps, one can expect that once all traps are filled, the remaining charge carriers are very mobile. Denote as $N_t$ the number of traps, with $N_c$ the number of charge carriers as in Section~\ref{sec:tof} and $\tau({N_c})$ the ToF depending on the number of charge carriers, evaluated, e.g., by Eqs.~\ref{eq:hitting_time} and~\ref{eq:tof2}. Then one expects the ratio $\tau(N_c)/ \tau(N_c+1)$ to be large for $N_c=N_t$. Analysis of this ratio then provides a qualitative indication of the \emph{effective number of traps} (depending on a definition of ''large'') in the transport network but not their location \change{or their physical nature}. In addition, the evaluation of $\tau(N_c)$ for large number of carriers is computationally cumbersome~\cite{chen_graph_2024}.

To avoid such expensive calculations and still gain insight into which molecules correspond to the $N_t$ traps, one can consider all different subsets $Q(N_t)$ of size $N_t$ of molecules in the network. We then consider the ToF of a single charge carrier $\tau(1)$ in the full system and in a system in which the subset $Q(N_t)$ has been removed from the network, $\tau^Q(1)$. The removal of the nodes is motivated by the fact that in the case of very strong traps, carriers will not easily escape them and thus they will for $N_c > N_t$ be largely inaccessible to the mobile carriers. In this scenario, one can inspect the ratio $\tau(1)/\tau^Q(1)$ and identify traps from its ''large'' values. \change{Such change in ToF provides a qualitative and system-agnostic way to identify traps in complex molecular charge transport networks, where site energy or other static properties alone may not suffice. }

\begin{figure}[tb]
    \centering
    \includegraphics[width=\linewidth]{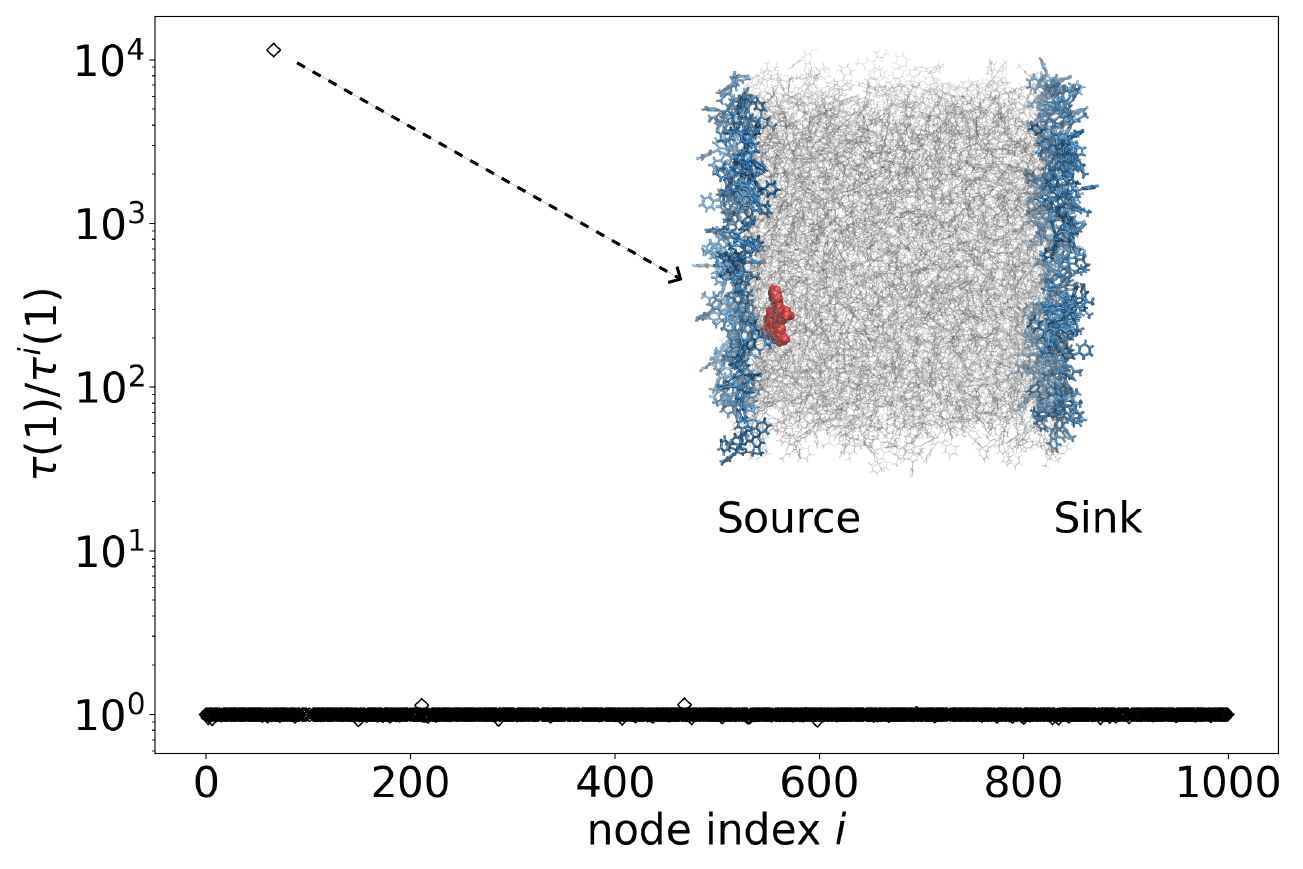}
    \caption{The scatter plot of $\tau(1) / \tau^i(1)$ for each node with index $i$. The only point with a value greater than 100 is the trap node whose index is 66. Inset: A box of 1000 BCP molecules simulated from MD. The Source and Sink molecules are highlighted by the blue color, and the molecule corresponding to trap node having large $\tau(1) / \tau^i(1)$ is highlighted by the red color. The other BCP molecules are in gray color.  }
    \label{fig:ToFFill}
\end{figure}

In the special case with only a single trap $N_t=1$ the set $Q(1)$ is given by a single node index $i$, and we can inspect $\tau(1)/\tau^i(1)$. This is shown for all $i$ for the simulated BCP system in Figure~\ref{fig:ToFFill}. It is clear that there is a single molecule node whose presence in the molecular charge transport network affects the ToF by four orders of magnitude. While it is intuitive to call this particular node a trap in this specific case, it is less obvious in general and the precise definition of a threshold value here is hardly possible. Another drawback of using the inspection of the ratio $\tau(1)/\tau^i(1)$, or its more general form, is the need for many cumbersome ToF calculations of the ${N \choose N_t}$ different scenarios. \change{It also fails to reveal any details about the physical nature of the trap(s).} This shows that a general method that identifies traps in molecular charge transport networks based on the graph structure alone without the need to actually calculate the dynamical properties is of great usefulness. \change{Its microscopic insight must, however, be compatible with the effects of traps on macroscopic observables such as the ToF here. Finally, we emphasize that due to the modifications of the BCP baseline system, and some modeling choices in the multiscale model, the reported charge transport properties, such as time-of-flight or mobilities, should not be compared to actual measured material properties, if available.} 

\subsection{Identification of Single-molecule Trap}
\label{sec:singletrap}
From the above analysis of the ToF sensitivity to the various nodes, we have identified node 66 is being a trap in the simulated BCP system, with site energy \unit[-1.89]{eV}. We will refer to this node as \emph{the trap node} and will denote it as $v_\mathrm{trap}$. In the following this trap node is used as the reference to scrutinize approaches for trap identification in molecular charge transport networks that solely rely on network properties.

\subsubsection{Results from Graph Decomposition Method}
\label{sec:GD}

\begin{figure}[tbp]
    \centering
    \includegraphics[width=\linewidth]{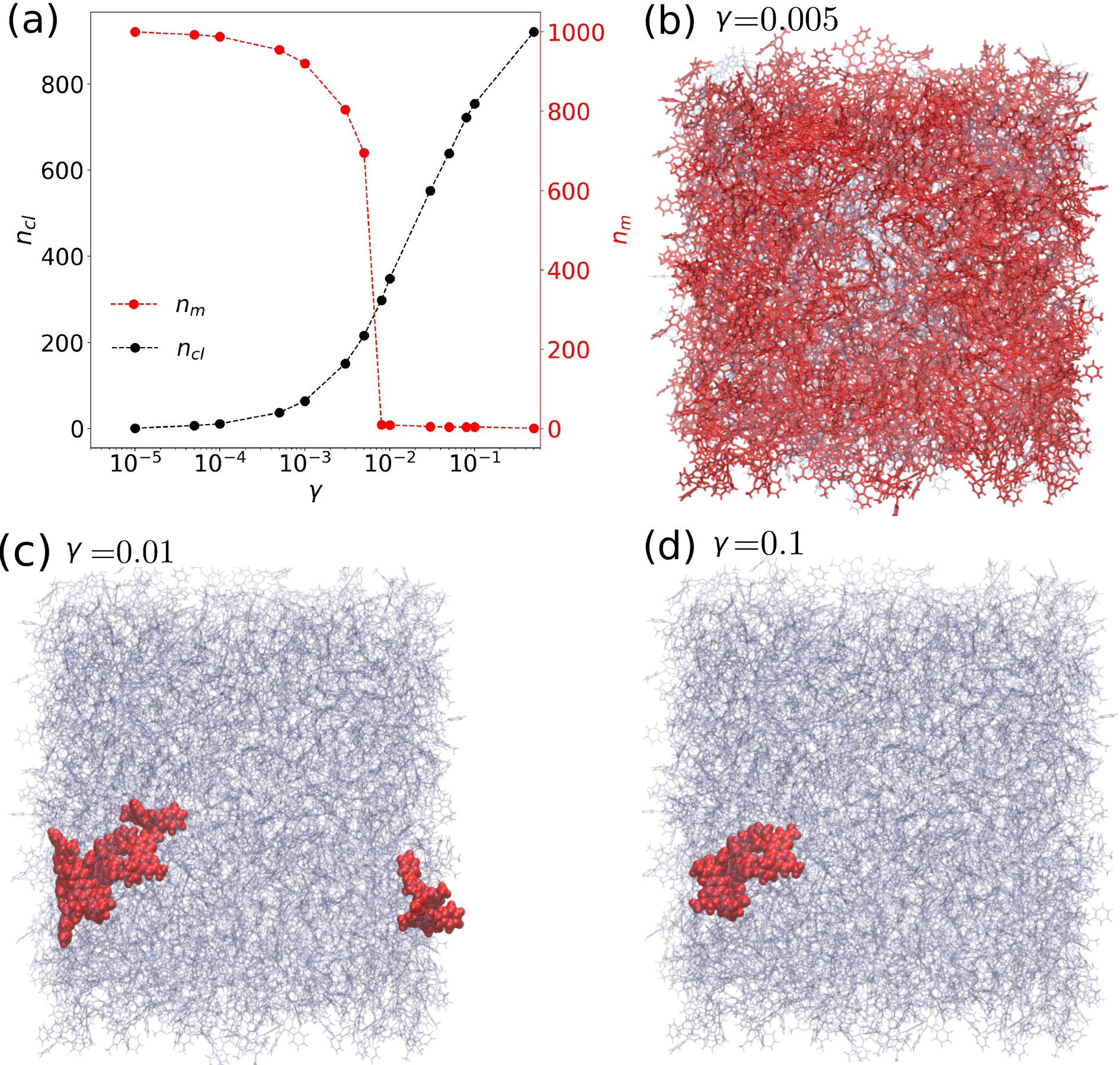}
    \caption{Results of GD method on the BCP charge transport network. (a) The total number of clusters ($n_\text{cl}$) after performing the GD method, and the size of the cluster containing the trap node ($n_\text{m}$), as a function of GD parameter $\gamma$. For three values of $\gamma$, the BCP structures are visualized: (b) $\gamma=0.005$, the red molecules indicate the cluster containing the trap node with more than 600 molecules. (c) $\gamma=0.01$, the 9 molecules in red indicate the cluster containing the trap node, and the molecules in blue color contain 348 clusters (d) $\gamma=0.1$ with 4 molecules in red forming the cluster containing the trap node.}
    \label{fig:GD}
\end{figure}

Before turning to the spectral clustering method as proposed in Section~\ref{subsec:SCT}, we first show the results of the graph-theoretical decomposition method~\cite{stenzel_general_2014}, whose technical details are summarized in~\ref{app:gd}. The method contains three adjustable parameters, $\alpha,\beta,\gamma$. The first two are related to details of the connectivity of the graph, whereas $\gamma$ is used to define a threshold involving the ratios of transition rates. In application to multi-node traps or trap regions, a choice of $\alpha,\beta = 0.02$ and $\gamma =0.2$ has been reported before~\cite{brereton_efficient_2014}. 

We apply the GD method to BCP and calculate the number of clusters $n_\text{cl}$ and the total number of molecules in the cluster containing the trap node, denoted as $n_\text{m}$ for different choices of the parameters. We find the results of the GD method to be mostly insensitive to the choice of $\alpha$ and $\beta$ as shown in Figure~\ref{fig:GD2} in~\ref{app:gd}. The parameter $\gamma$ related to the rate ratios has, however, a significant impact on the obtained clustering, as is shown in Figure~\ref{fig:GD}(a) for the range of  $10^{-5} \leq \gamma \leq 2 \cdot 10^{-1}$. For very small $\gamma$, we obtain a single large cluster equal to the whole system. With increasing $\gamma$ the GD method yields more clusters. Between $\gamma=10^{-3}$ and $10^{-2}$ one can see a rather sharp transition in the size of the cluster that contains the previously identified trap node. At the onset of the transition ($\gamma=5\cdot10^{-3}$), see Figure~\ref{fig:GD}(b), the cluster containing this node contains more than 600 molecules. After the transition ($\gamma=10^{-2}$, Figure~\ref{fig:GD}(c)), this size is massively reduced. For the previously recommended value of $\gamma=0.2$, each molecule is a cluster by itself. Tuning the parameter values of $\alpha, \beta$ does not help in identifying the trap node and GD cannot directly detect the single trap node that leads to large ToF, although the GD method should group molecules into clusters where random walk jumps are more frequent compared to jumps outside the clusters. 

\subsubsection{Results from Spectral Clustering Method} 

\begin{figure*}[tbp]
    \centering
    \includegraphics[width=\linewidth]{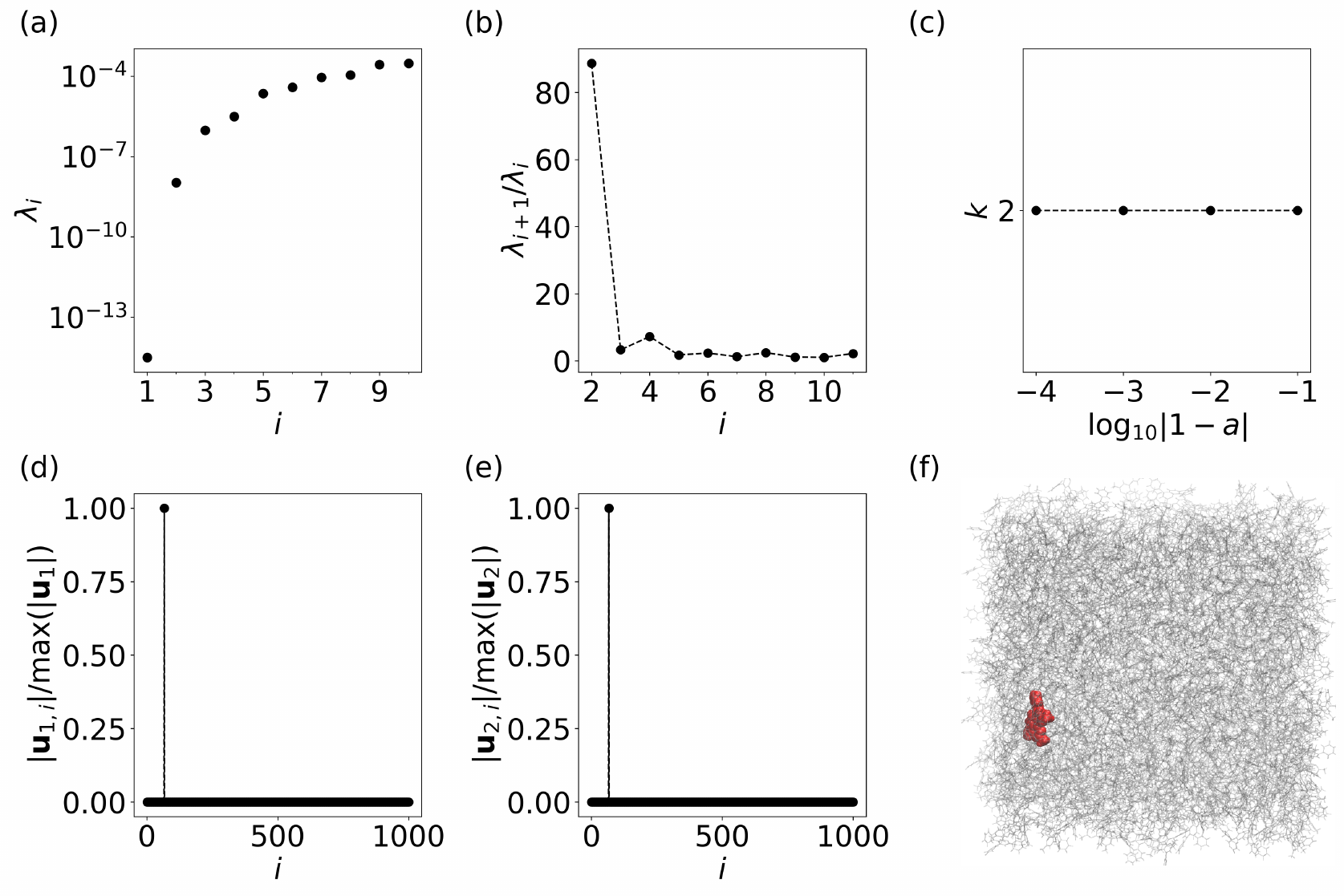}
    \caption{Results of the spectral clustering method for multiscale modeled BCP system: (a) The first ten eigenvalues of $L_\text{rw}$. (b) $\lambda_{i+1}/{\lambda_i}$ as a function of $i$. (c) The number of clusters determined from Algorithm \ref{alg1} as a function of $a$. (d)-(e) The first and second normalized eigenvector elements of $L_\text{rw}$ as a function of node indexes. (f) The BCP system where molecules in red and grey color are first and second clusters, respectively, as identified using the K-means clustering with $k=2$.}
    \label{fig:66RW}
\end{figure*}

Turning to the performance of the spectral clustering method in application to BCP, we first show in Figure~\ref{fig:66RW}(a) the ten smallest eigenvalues of the random-walk Laplacian. We note that the first eigenvalue is numerically very close to zero, as expected by the theory. Focusing on the next two eigenvalues, one can see a strong increase from $\lambda_2$ to $\lambda_3$, while the preceding eigenvalues increase less quickly. These observations lead to a behavior of the ratio $\lambda_{i+1}/\lambda_i$ as shown in Figure~\ref{fig:66RW}(b) for $2 \leq i \leq 10$. The maximum of $\lambda_{i+1}/\lambda_i$ is larger than 80 when $i=2$, while all other values are below 5. Our procedure in Algorithm~\ref{alg1} will thus take $\ell = 2$ in line 5. Figure~\ref{fig:66RW}(c) shows the outcome of the algorithm for different values of the threshold parameter $a$ ranging from \unit[0.9]{} to \unit[0.9999]{}. We note it robustly yields the value $k=2$ for the K-means clustering step, meaning that there should be one trap in the system. The reason for the observed robustness is clear from the inspection of the normalized elements of the first and second eigenvectors in Figure~\ref{fig:66RW}(d) and (e), respectively, which show a single entry (corresponding to the trap node) being 1. Accordingly, the K-means clustering step is performed with $k=2$  yielding one cluster that only contains the previously identified trap node, while the other cluster is the rest of the system, visually indicated in Figure~\ref{fig:66RW}(f). 

\subsubsection{Relation between Cost function and ToF}

\begin{figure}[tbp]
    \centering
    \includegraphics[width=\linewidth]{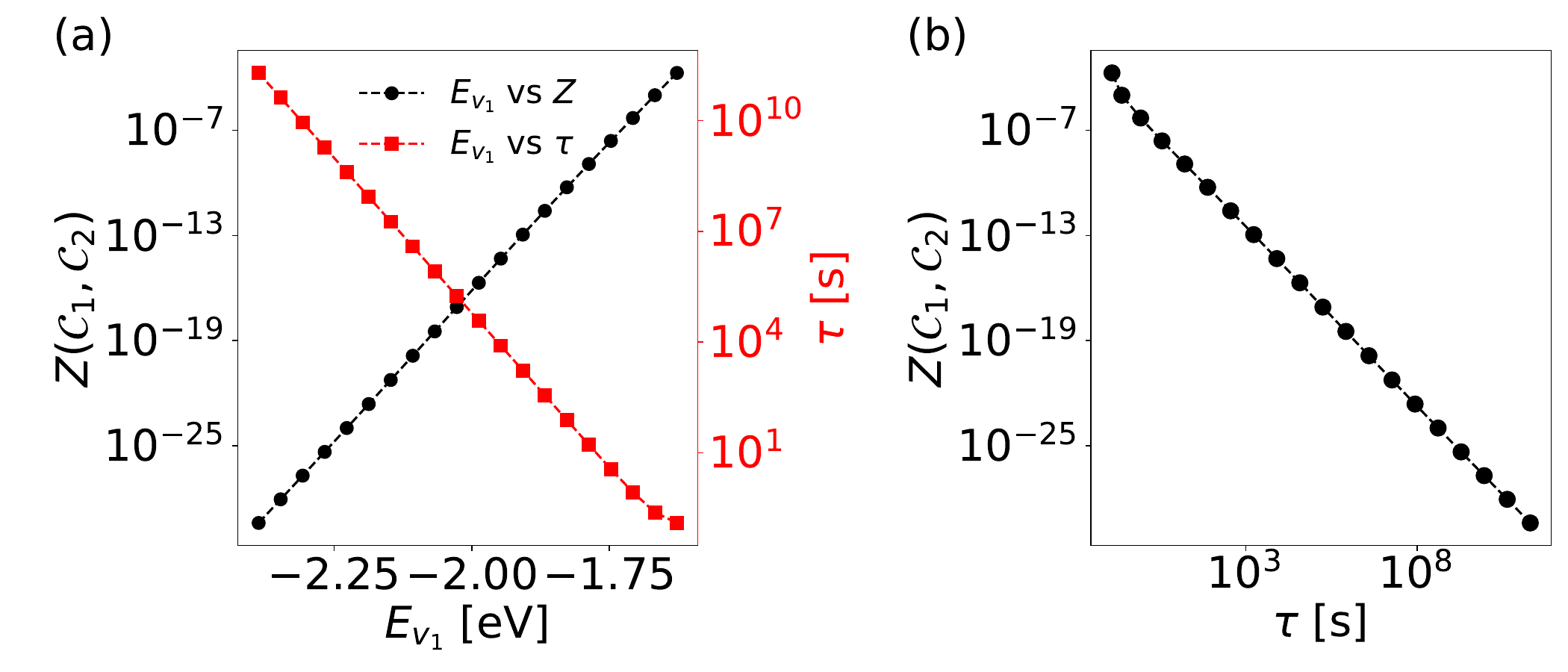}
    \caption{In a BCP system, the trap node's energy $E_\text{trap}$ is varied from \unit[-2.4]{eV} to \unit[-1.6]{eV}. (a) Dependence of the cutting cost $Z(\mathcal{C}_1,\mathcal{C}_2)$ and the ToF $\tau$ on $E_\text{trap}$. (b) Dependence of $Z(\mathcal{C}_1,\mathcal{C}_2)$ on $\tau$.}
    \label{fig:E_ToF_Z_nonsym}
\end{figure} 

The spectral clustering method very clearly identified the trap node in the BCP system. This is also reflected by the fact that the value of the cost function~\eqref{eq:cost} for the proposed cut is $8.8 \cdot 10^{-12}$, while the cost of cutting any other single node is $0.99$. We thus see a clear indication that a low cost is a qualitative signal of a possible trap. On the other hand, it is known that the energy of a trap node qualitatively influences the recorded ToF. This raises whether there is also a relation between the site energy of single-molecule traps and the cost of the corresponding cut, and hence between the cost function and the ToF. To investigate this, we take the simulated BCP system, modify the site energy $E_\text{trap}$ of the trap node, ranging from a small energy value (\unit[-2.4]{eV}) to a relatively large value (\unit[-1.6]{eV}) and determine the cost of cutting according to our method this node and the ToF in the system. We note that when the original trap node has a site energy above (\unit[-1.6]{eV}) it is no longer defined as a trap by our algorithm, which explains why we use this as the upper bound on the energy for this experiment.

The results are shown in Figure~\ref{fig:E_ToF_Z_nonsym}(a). For low values of the site energy, one simultaneously observes a large ToF and  small cost, indicating both a clear trap characteristic and straightforward identification via our spectral clustering method. Increasing the energy leads to a decrease of ToF and an increase of the cost of cutting the node from the graph. Moreover, when plotting the pairs of cost function as a function of the ToF, as in Figure~\ref{fig:E_ToF_Z_nonsym}(b), we observe a power-law dependence between the two quantities. Together these results show that indeed the cost function associated with cutting out a single-molecule trap and the ToF of the system are intricately related. 

\change{
\subsubsection{Effect of an Applied Electric Field}

\begin{figure}[tbp]
  \centering
  \includegraphics[width=0.9\linewidth]{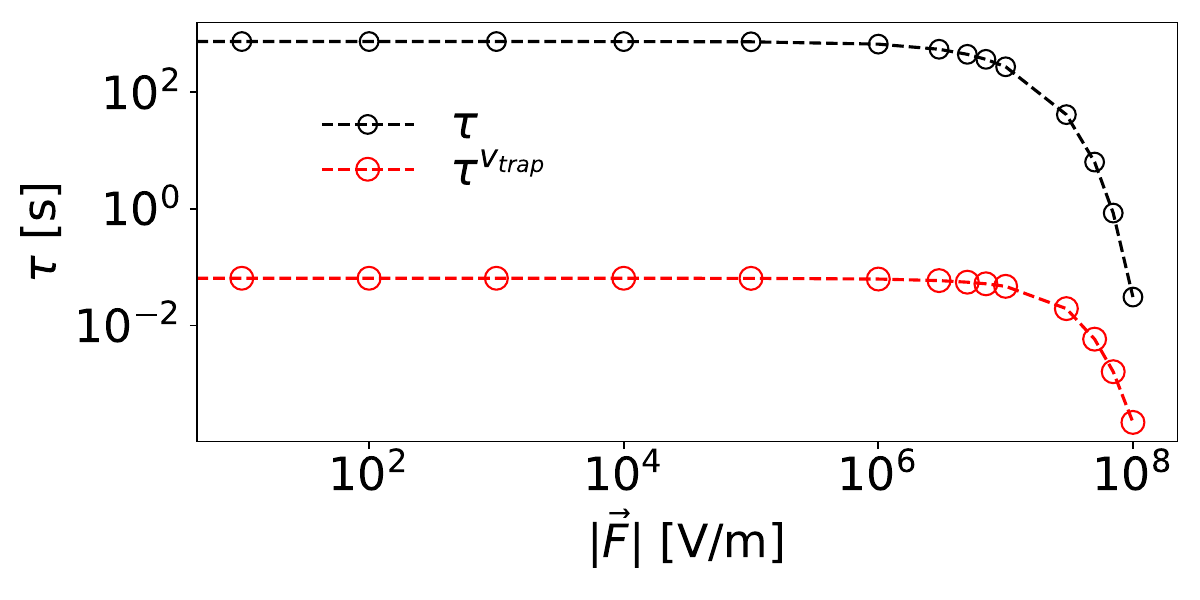}
  \caption{(a) The 1-carrier ToF $\tau$, and the ToF $\tau^{v_\text{trap}}$ when removing the node $v_\text{trap}$ as a function of the electric field.}
  \label{fig:ElectricField}
\end{figure} 

To further validate the robustness of our spectral clustering method, we investigate its performance under non-equilibrium conditions by introducing an external electric field as typically used in operating conditions for devices based on organic semiconductors. The charge dynamic is then a drift-diffusion process rather than a diffusion. The electric field we applied is $\vec{F} = [F_x , 0 , 0]$ where $F_x$ has values in the range $0 $ to $ \unit[10^{8}]{V/m} $. 

The inclusion of an electric field in the Marcus rates modifies the transition rates between molecules, as the field introduces a directional bias in charge carrier motion.
Figure~\ref{fig:ElectricField} shows that as the electric field strength increases above $\unit[10^6]{V/m}$, the ToF decreases significantly, indicating a faster charge transport along the direction of $\vec{F}$. Using the proposed method, the node $v_\text{trap}$ is found as a single cluster across the range of the studied electric field. Removing $v_\text{trap}$ results in a significant decrease in ToF as indicated by the gap between $\tau$ and $\tau^{v_\text{trap}}$. As the electric field $|\vec{F}| > \unit[10^{6}]{V/m}$, the gap between $\tau$ and $\tau^{v_\text{trap}}$ shows signs of reducing. This phenomenon suggests that an electric field, having the direction drift force for the charge dynamics, can potentially reduce the trap effects.}

\subsection{Identification of Multiple Distributed Traps}

\begin{figure}[tbp]
    \centering
    \includegraphics[width=\linewidth]{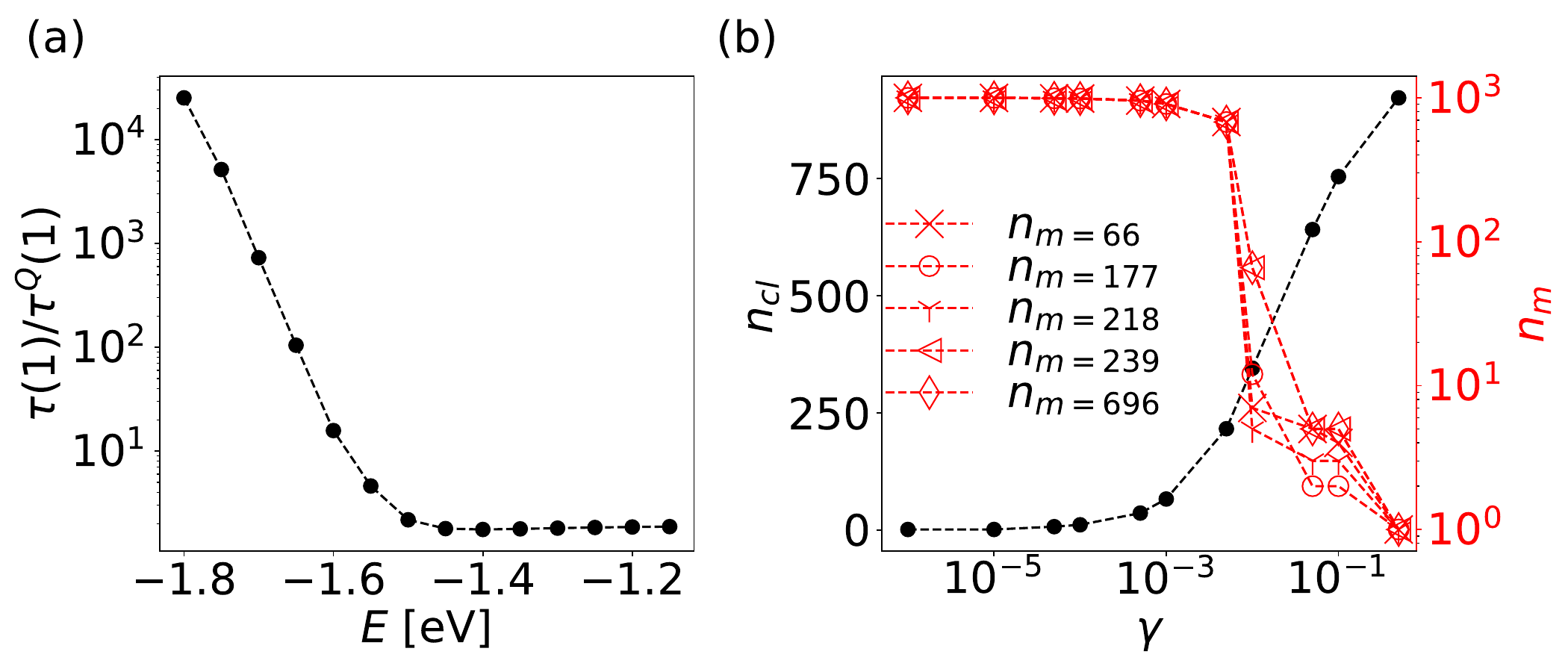}
    \caption{Sensitivity of ToF and performance of the GD method for multiple distributed traps. (a) The ratio $\tau(1)/\tau^Q(1)$ as a function of the energies of the nodes in $ Q = \{ v_1, v_2, \cdots, v_5 \}$. (b) The number of clusters $n_\text{cl}$ after performing the GD method when the nodes in $ Q $ have energies $E=\unit[-1.8]{eV}$, and the size of the cluster containing a specific node $n_{m_i}$, as a function of GD parameter $\gamma$.}
    \label{fig:GD_5trap}
\end{figure} 

The previous section shows that the spectral clustering method can reliably identify a single trap node as it occurs in the modeled BCP system. In the following, we will scrutinize if the same holds in a system with multiple trap molecules which are not connected with each other --  a scenario we refer to as {\em multiple distributed traps}. Within the spectral clustering method, we expect that having multiple distributed traps will increase the cluster number $k$ for the K-means clustering steps as given by Algorithm~\ref{alg2}. As the simulated BCP system does not have multiple distributed traps, we modify it taking the trap node we found and 4 other nodes ($Q = \{v_1=v_\mathrm{trap}, v_2 \cdots, v_5\}$), which are not connected by an edge in the charge transport network and set their site energies all to $E=\unit[-1.8]{eV}$, which is close to the site energy of the original trap node. It can be seen from the dependence of the ratio $\tau(1)/\tau^Q(1)$ on $E$ in Figure~\ref{fig:GD_5trap}(a) that such a value points indeed to a very pronounced trapping effect of the charge carrier. 

First, we consider the predictions from the GD method in Figure~\ref{fig:GD_5trap}(b). We again show the resulting number of clusters $n_\text{cl}$ and the number of molecules $n_{m_i}$ in each of the clusters containing one of the 5 prepared trap nodes $v_i$, depending on the parameter $\gamma$ in the GD method. Qualitatively, one can observe the same behavior as for the single trap case: the method only yields isolated traps if nearly each molecule is its own cluster, or the whole system is a single cluster. Evidently, the method fails to correctly characterize the multiple distributed traps situation in the charge transport network.   

\begin{figure*}[tbp]
    \centering
    \includegraphics[width=\linewidth]{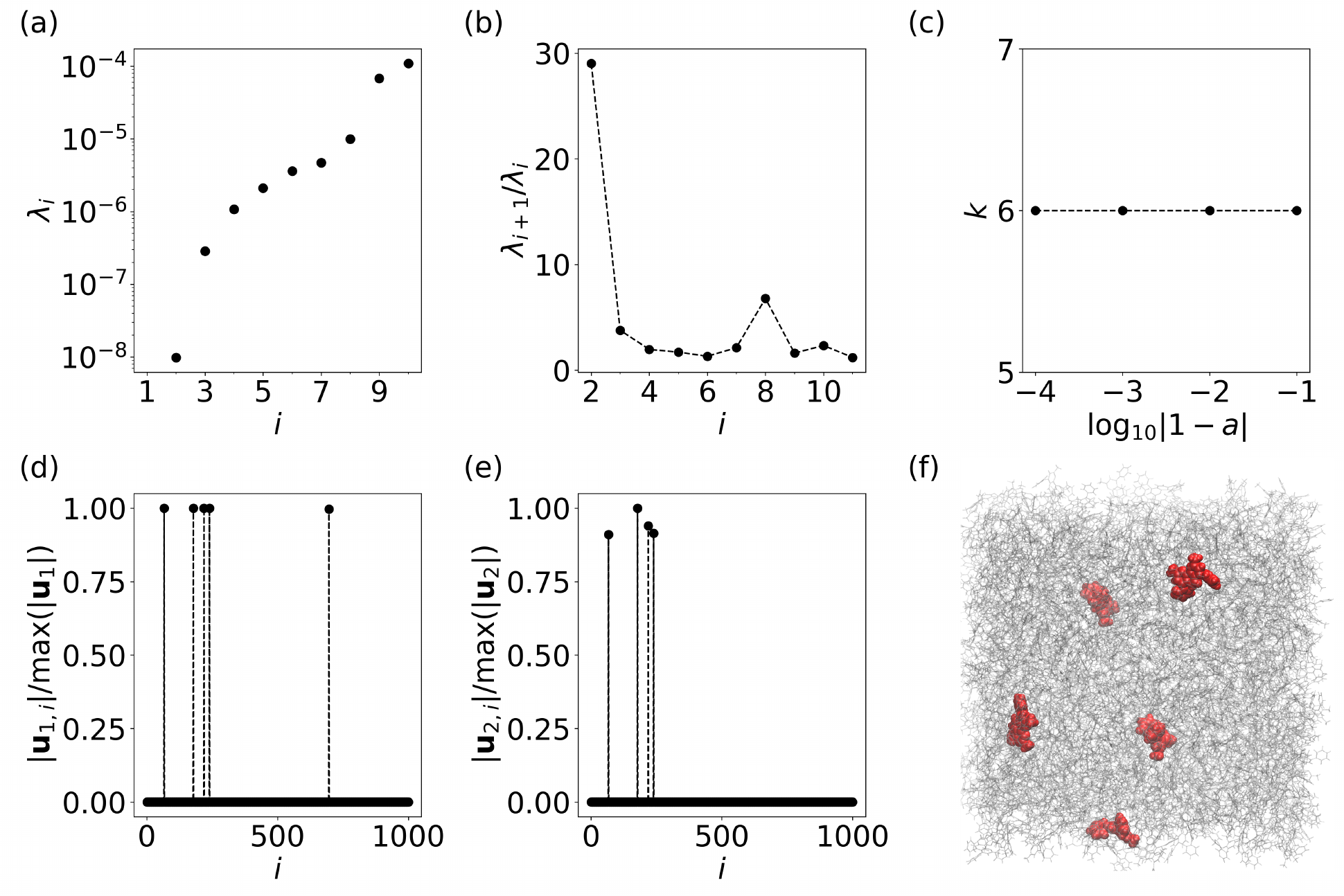}
    \caption{Results of the trap identification by spectral clustering methods for the BCP systems with multiple distributed traps. (a) The first ten eigenvalues of $L_\text{rw}$. (b) ${\lambda_{i+1}}/{\lambda_i}$ as a function of $i$. (c) The number of clusters determined from Algorithm \ref{alg1} as a function of $a$. (d)-(e) The first and second eigenvector elements of $L_\text{rw}$ as a function of node indices. (f) The BCP system where molecules in red color are molecules consisting of nodes $v_1, v_2, \cdots, v_5$. Each red molecule is partitioned as one cluster. The grey molecules are one cluster, as identified using the K-means clustering with $k=6$.}
    \label{fig:fig_5trap}
\end{figure*} 

Turning to our spectral clustering method, we see in Figure~\ref{fig:fig_5trap}(a) and (b) that the largest value for the ratio is still at $i = 2$, as was the case for the single trap situation. However, unlike that situation, 
when inspecting the first two normalized eigenvectors in Figure~\ref{fig:fig_5trap}(d,e), one can clearly see five and four relevant non-zero elements, respectively. Overall, the determined number of unique induced subgraphs is $k=6$, which indeed implies that there are 5 traps in the system shown in Figure~\ref{fig:fig_5trap}(f): the five individual distributed trap nodes and one cluster that is the rest of the system. This shows the necessity of the addition step in Algorithm~\ref{alg2}. The result $k = 6$ is also relatively robustly with respect to the choice of the method's parameter $a$ as shown in Figure~\ref{fig:fig_5trap}(c). Only when $a>0.9$ is used, the spectral clustering method yields $k=2$. So again, using Algorithm~\ref{alg1} with $a=0.9$ finds the correct number of clusters $k=6$, and also the right trap nodes.

\subsection{Identification of Trap Region}

\begin{figure}[tbp]
    \centering
    \includegraphics[width=\linewidth]{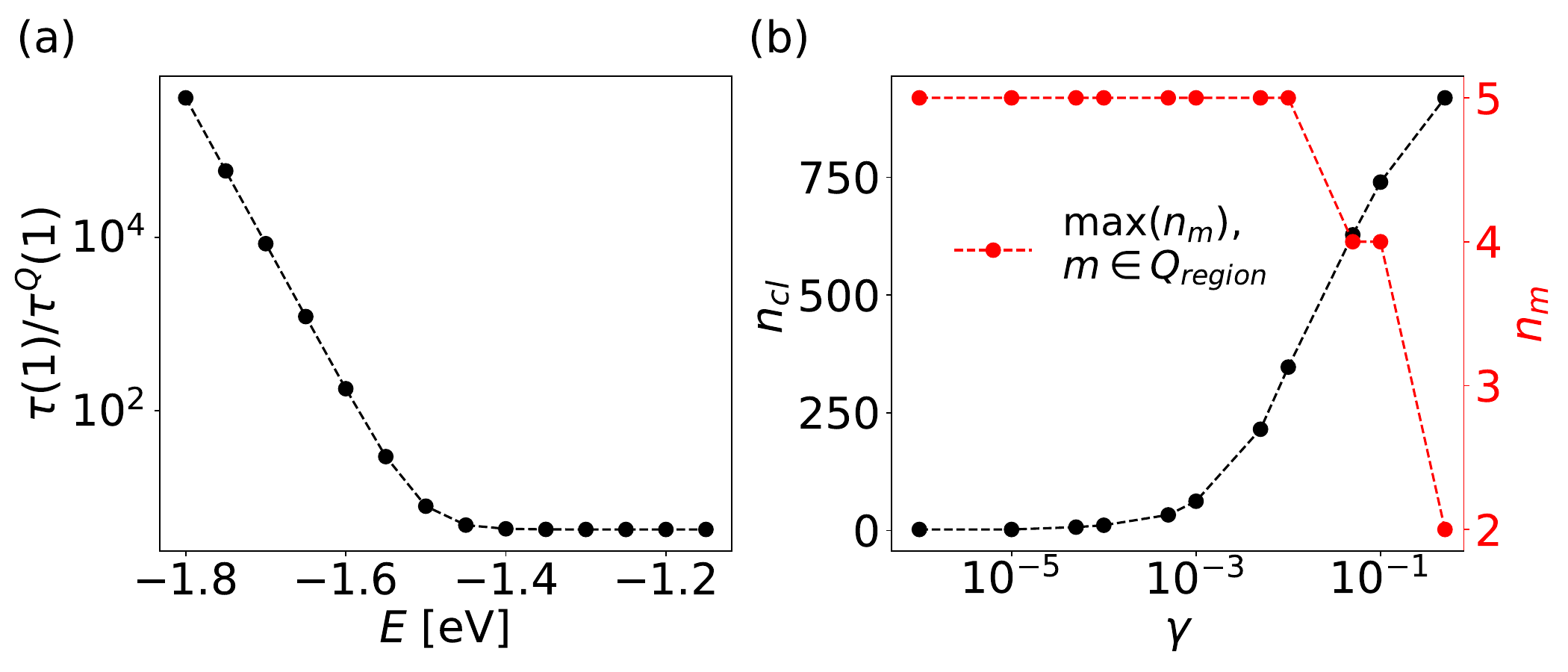}
    \caption{Sensitivity of ToF and performance of the GD method for trap regions. (a) The ratio $\tau(1)/\tau^Q(1)$ as a function of the energies of the nodes in $Q = \{v_1, v_2, \cdots,v_6\}$. (b): The number of clusters after performing the GD method when the nodes in $Q$ have energies $E=\unit[-1.8]{eV}$, and the maximum size of the cluster containing at least one specific node in $Q$, as a function of GD parameter $\gamma$.}
    \label{fig:GD_696region}
\end{figure} 

\begin{figure*}[tbp]
    \centering
    \includegraphics[width=\linewidth]{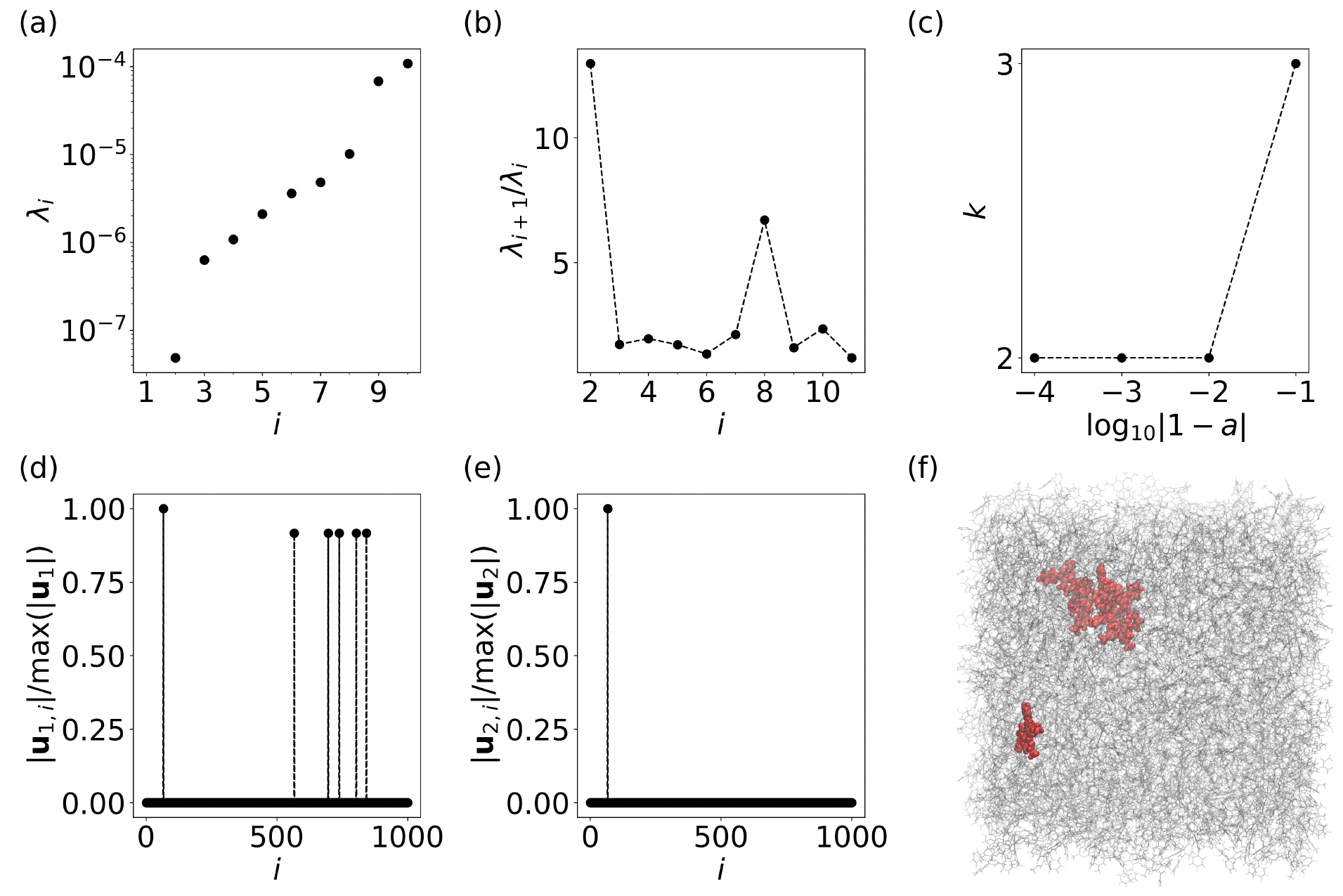}
    \caption{Trap identification results using spectral clustering methods on the BCP system with trap regions. (a) The first ten eigenvalues of $L_\text{rw}$. (b) $\lambda_{i+1}/{\lambda_i}$ as a function of $i$. (c) The number of clusters determined from Algorithm \ref{alg1} as a function of $a$. (d)-(e) The first and second eigenvector elements of $L_\text{rw}$ as a function of node indices. (f) The BCP system where molecules in red color show $v_1$ as a single molecule cluster and the connected region $Q = \{ v_2, \cdots,v_6 \}$ is a second cluster. The gray molecules form the third cluster, as identified using  K-means clustering with $k=3$.}
    \label{fig:region696}
\end{figure*} 

It is known that for some materials the site energies are correlated in space, e.g., due to strong permanent dipole interactions as in amorphous Alq$_3$~\cite{Baumeier2011,baumeier_stochastic_2012,brereton_efficient_2014}. This leads to a situation in which several sites with low energy are connected, forming a trap region instead of single-molecule traps as studied in the previous two sections. A random walk process that enters such a region tends to jump among those nodes with only a small probability of escaping the region. This scenario was the original motivation for the GD method as it aims at partitioning the graph into clusters that the random walk spends a significant amount of time in.

Here we will investigate whether our proposed spectral clustering methods is able to identify such trap regions in the BCP system. Because the BCP system does not have strong spatial correlations, there are no such trap regions in the as-modeled molecular charge transport network. We therefore select six nodes $Q = \{v_1=v_\mathrm{trap}, v_2, \dots, v_6\}$ consisting of the original trap node and a distant node $v_2$ disconnected from $v_1=v_\mathrm{trap}$ plus its four closest neighbors $v_3,v_4,v_5,v_6$. We then set the site energies of the nodes in $Q$ to a value $E$ so the $v_\mathrm{trap}$ is a single-molecule trap while $Q_\text{region} := \{v_2, \dots, v_6\}$ is trap region. In Figure~\ref{fig:GD_696region}(a), we show $\tau(1)/\tau^Q(1)$ for different values of $E$ and observe a clear trapping effect on the ToF for $E<\unit[-1.6]{eV}$. In the following we set $E=\unit[-1.8]{eV}$.

Figure~\ref{fig:GD_696region}(b) shows the result of the GD method to identify the traps: the total number of clusters $n_\text{cl}$ and the maximum cluster size $n_m$ when $m$ is one of the nodes in $Q_\text{region}$ for different values of $\gamma$. When $\gamma=10^{-6}$, $n_m=5$, and $Q_\text{region}$ is successfully identified as a single cluster and the rest of the 995 nodes as another cluster. As in Section~\ref{sec:singletrap}, the isolated trap node is not identified as a separate cluster. As $\gamma$ increases, more clusters appear, and when $\gamma > 10^{-2}$, the five low-energy nodes in $Q_\text{region}$ are no longer partitioned into one cluster.

Turning now to the trap identification by spectral clustering, we show the eigenvalues of $\mathbf{L}_\text{rw}$ and the ratio $\lambda_{i+1}/\lambda_i$ in Figure~\ref{fig:region696}(a,b), respectively. We again observe that the maximum of the ratio is found at $i=2$. The full algorithm yields a value $k=3$, indicating the presence of two traps. As can be seen from Figure~\ref{fig:region696}(c), the determination of $k$ is very robust with respect to the choice of the parameter $a$. Looking at the elements of the first two normalized eigenvectors (see Figure~\ref{fig:region696}(d,e)) we clearly see a single large entry and a group of five other large entries, representing $Q_\text{region}$. Finally, K-means clustering with $k=3$ yields the three correct clusters as shown in Figure~\ref{fig:region696}(f): in red the isolated trap node $v_\mathrm{trap}$ and the trap region formed by $Q_\text{region}$ and in gray the third cluster containing all other molecules.

It should be noted that there is a second large jump for the eigenvalues at $i = 4$. To further check the performance of our method we also perform the spectral clustering method on the first four eigenvectors, instead of the first two. This gives $k=4$ with the only difference being that there is now a fourth cluster which comprises a single node $v_7$. However, when checking the effect of this node on the ToF via the ratio $\tau(1)/\tau^Q(1)$ with node $v_7$ added to $Q$, we find that the ratio is $\approx 1$, i.e., node $v_7$ does not have the system characteristics we associated with a trap. This supports the choice of $\ell=2$ yielded by our Algorithm~\ref{alg1}. 

\change{

\begin{figure}[tbp]
  \centering
  \includegraphics[width=\linewidth]{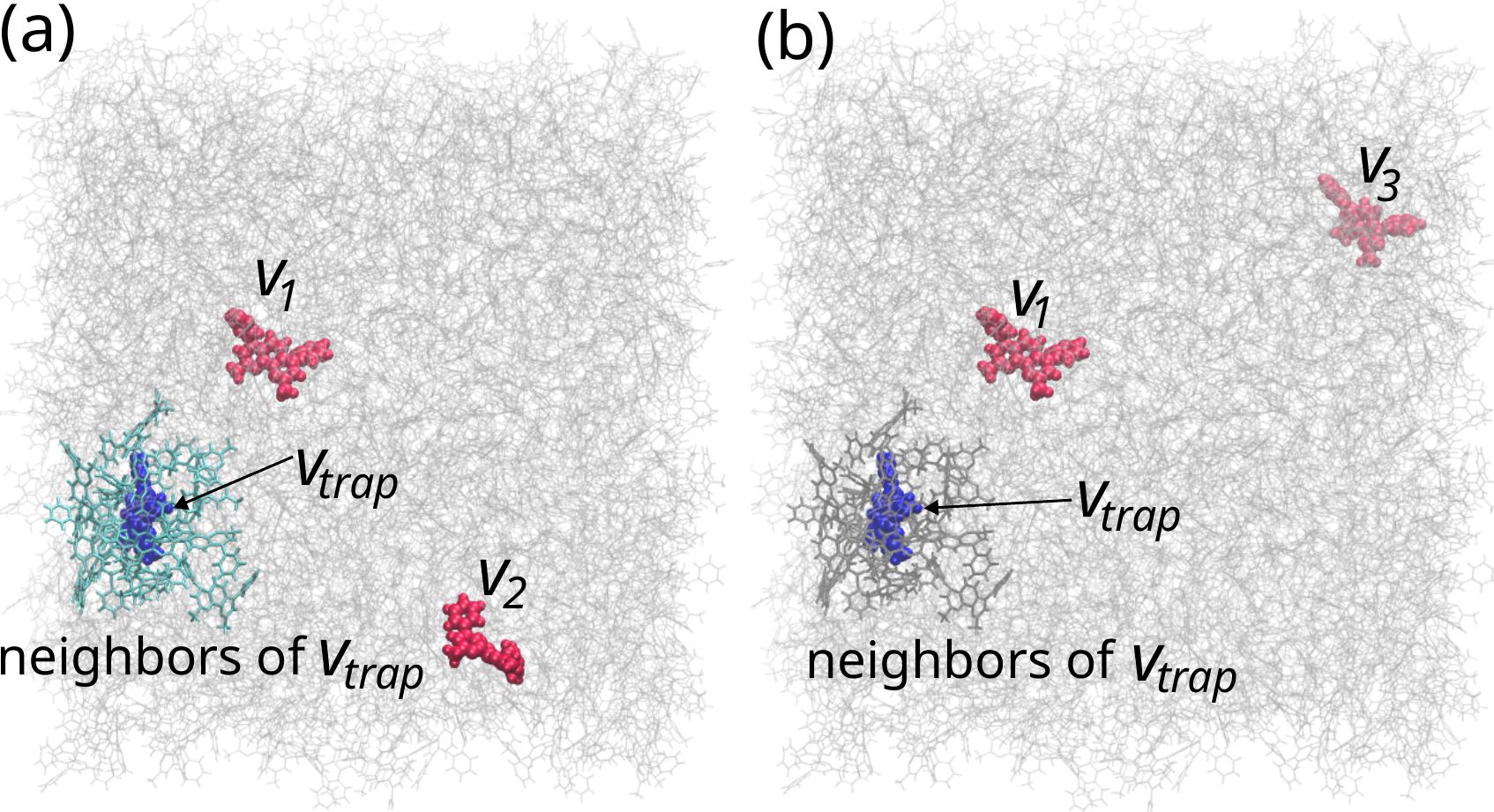}
  \caption{Visualization of two non-energy based trapping scenarios: (a) the node $v_\text{trap}$ (blue) with the lowest energy and its (green) represent different molecular types with non-symmetric reorganization energies is correctly identfied as not being a trap. Two BCP molecules in red color, $v_1$ and $v_2$, are two single clusters identified as actual traps using the K-means clustering with $k=3$. (b) The neighbors of $v_\text{trap}$ are colored in light-dark and their electronic coupling element with $v_\text{trap}$ are small. Two BCP molecules in red color, $v_1$ and $v_3$, are two single traps identified using the K-means clustering with $k=3$ in this topology-based scenario.}
  \label{fig:2vmds}
\end{figure}

\subsection{Non-symmetric reorganization energies}
In materials containing different types of molecules and complex molecular geometries, the reorganization energy $R_{ij}$ for charge transfer between molecules is non-symmetric, i.e., $R_{ij}\neq R_{ji}$. In such cases, the rates for this transfer do not obey detailed balance and the equilibrium Boltzmann population argument mentioned in the Introduction cannot be applied, and the lowest energy site may not necessarily act as a trap, even if it has a significantly lower site energy compared to the rest of the network.

To illustrate that our proposed method is also able to correctly identify traps in such a situation, we modify the BCP system by introducing a non-symmetric reorganization energy for specific pairs of molecules. We select the node $v_\text{trap}$ with the lowest site energy $E_{v_\text{trap}}= \unit[-1.80]{eV}$ and adjust the reorganization energy $R$ such that $R_{v_\text{trap} \, j} = \unit[0.1]{eV}$ and $R_{j \, v_\text{trap}} = \unit[0.2]{eV}$. We show this setup in Figure~\ref{fig:2vmds}(a), where the node $v_\text{trap}$ is colored in blue and its neighbors are in lighter color.

Due to the non-symmetric reorganization energy, the transition rate from $v_\text{trap}$ to its neighbors is not small, despite the low energy. This results in a situation where the low-energy site is not a trap because charge carriers can easily escape from it due to the high forward transition rate. In contrast, the spectral clustering method (see detailed results in Fig.~\ref{fig:lambdaVary} in the Appendix), which considers the full graph structure and transition rates, correctly identifies that this site is not a trap. Instead, the  method identifies nodes $v_1$ and $v_2$ with site energy $E_{v_1} = E_{v_2} =  \unit[-1.70]{eV}$ (in red color shown in Fig.~\ref{fig:2vmds}(a)) as the isolated single traps in this system, putting those two nodes in their own cluster and the other 998 nodes in another. 

The actual calculated ToF support the results of the spectral clustering method. For the system without removing any node we find $\tau=\unit[5.7\cdot 10^{-1}]{s}$. Removing $v_\text{trap}$, one obtains  $\tau^{v_\text{trap}}=\unit[5.5\cdot 10^{-1}]{s}$, while removing $v_1, v_2$  yields $\tau^{(v_1, v_2)}=\unit[1.4 \times 10^{-3}]{s}$. We clearly see that removing the lowest energy site $v_\text{trap}$ has hardly any effect on the ToF, while the nodes identified from our proposed method influence the ToF by more than two orders of magnitude.

\subsection{Topological coupling-based traps}
Next to mixed molecular material compositions, in general structural features can also lead to charge trapping behavior in materials. To mimic such a scenario within the as-modeled BCP baseline system, we modify the electronic coupling element $J_{v_\text{trap} \, j}$ of the lowest-energy site $v_\text{trap}$ with its neighboring sites $j$. Specifically, the electronic coupling elements $J_{v_\text{trap} \, j}$ between the lowest energy site $v_\text{trap}$ (highlighted in blue in Fig.~\ref{fig:Jshield}) and its neighbors $j$ has range: $\unit[10^{-9} ]{eV^2} > J_{v_\text{trap} \, j}^2 > \unit[10^{-14}]{eV^2}$. Such a system is visualized in Figure \ref{fig:2vmds}(b), where the lowest energy site $v_\text{trap}$ is in color blue and its neighbors are in light-dark color.

In such a case, even if a site has a very low energy, the probability of charge carriers transitioning to or from this site is extremely low, resulting in low occupancy and hence no trapping effect in the non-steady state. While using a solely energy-based argument would still classify $v_\text{trap}$ as a trap due to its low energy, it does not exhibit any trapping effect in practice, as carriers are unlikely to be captured or released from this site. The ToF of this system is $\tau=\unit[6.0\cdot 10^{-1}]{s}$, which is not chnaged after removing $v_\text{trap}$. Applying our proposed spectral clustering method (see the full results in Fig.~\ref{fig:Jshield} in the Appendix), 
gives the same $v_1$ as before and a new site $v_3$ as two single clusters while the remaining 998 nodes form a single large cluster. Removing $v_1, v_3$, the ToF is $\tau^{(v_1, v_3)}=\unit[6.3 \cdot 10^{-2}]{s}$, indicating the real trapping effect of $v_1, v_3$.}

\section{Conclusion}
\label{sec:con}
\change{In this study, we have developed and implemented a novel method based on spectral clustering to identify traps of widely different characteristics within molecular charge transport networks. Our approach is based on the application of K-means clustering on the eigenvector elements from the Laplacian matrix with a proposed heuristic based on eigenvalue ratio and eigenvector entries. The method avoids the complexities associated with calculating multiple-carrier ToF and overcomes the need for system-dependent parameters. Its effectiveness is demonstrated through the analysis of several trap types prepared with a baseline multiscale model. We successfully identified energy-based traps with a single-molecule, distributed molecule, and region characteristics. Our method also yields the correct identification of topology-based traps and is applicable to other, more general, situations (e.g., in mixtures with non-symmetric reorganization energies) in which the transition rates do not obey detailed balance.

This capability to identify and quantify traps without extensive parameter tuning marks a significant advancement over other candidate methods such as the watershed algorithm and graph-theoretic decomposition, or purely energy-based arguments. The added value of the spectral clustering method lies in its ability to capture the interplay between site energies, transition rates, and the overall network structure. This makes it a more robust and general approach for identifying traps in organic semiconductors, particularly in systems with complex charge transport dynamics.}

\section*{Acknowledgments}
BB acknowledges support by the Innovational Research Incentives Scheme Vidi of the Netherlands Organisation for Scientific Research (NWO) with project number 723.016.002. ZC and BB also are grateful for funding and support from ICMS via project MPIPICMS2019001.

\appendix 
\section{Electronic Structure Calculation}
\label{app:es}
\begin{figure*}[h]
    \centering
    \includegraphics[width=0.99\textwidth]{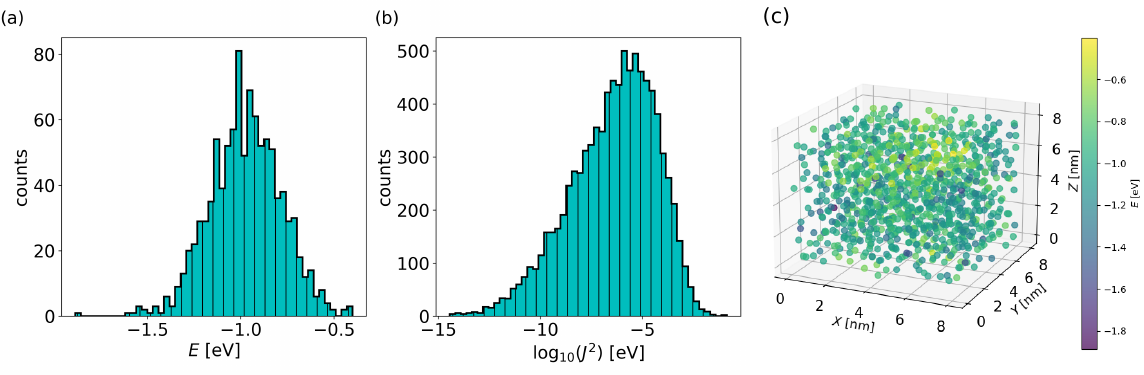}
    \caption{(a) The distribution of site energies. (b) The distribution of coupling elements. (c) The spatial resolution of molecule energies in 3-dimension, where each solid circle represents a molecule.}
    \label{fig:E_J}
\end{figure*}

The electronic structure parameters of the BCP molecule calculated from the multiscale model are shown in Figure\ref{fig:E_J}. 

\section{Graph Decomposition Method}
\label{app:gd}
\begin{figure*}[h]
    \centering
    \includegraphics[width=0.8\textwidth]{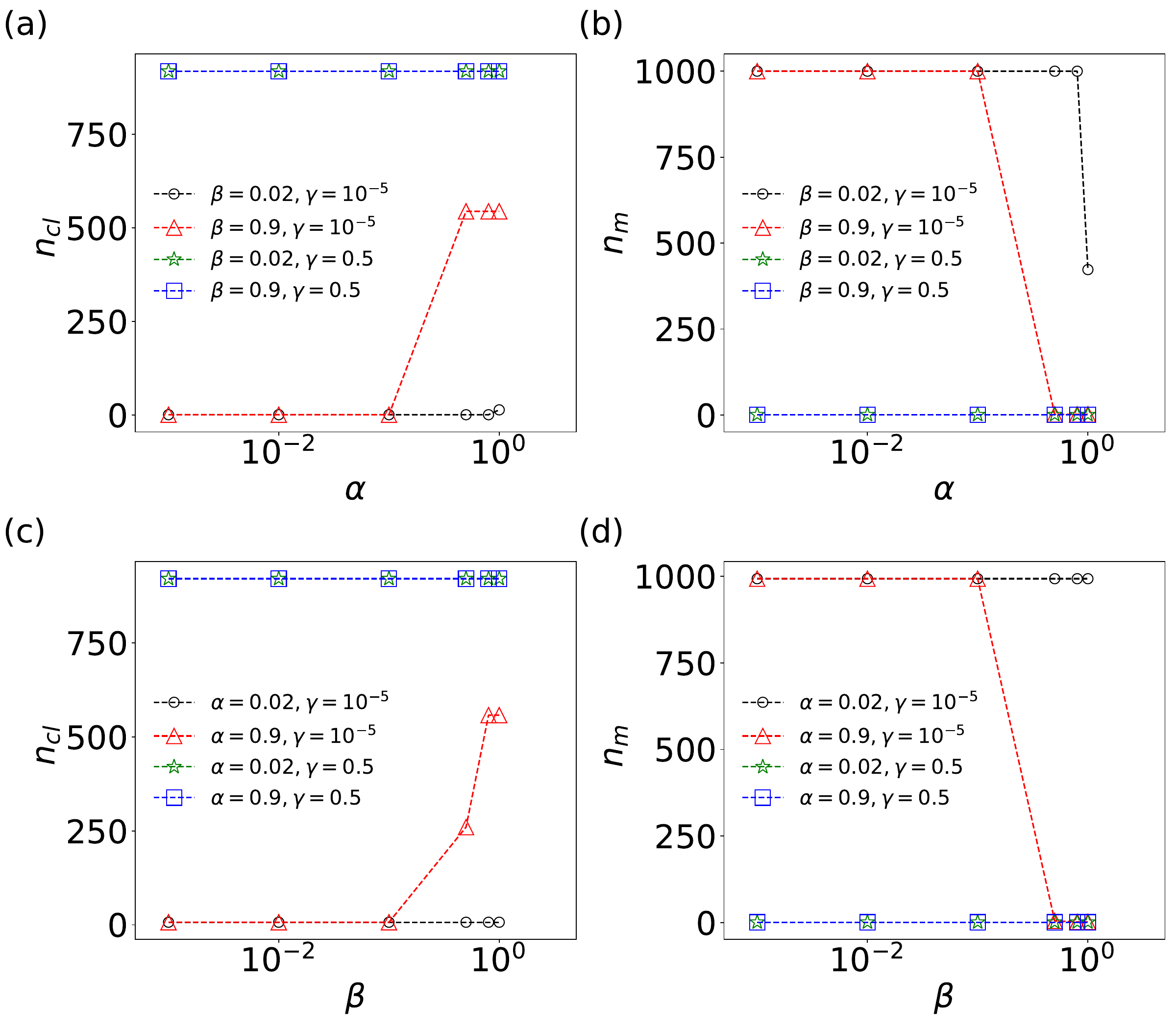}
    \caption{(a) The number of clusters after performing the GD method and (b) the size of the cluster containing the trap node with $\beta=0.02,0.9$ and $\gamma=0.5,10^{-5}$, as a function of $\alpha$. (c) The number of clusters after performing the GD method and (d) the size of the cluster containing the trap node with $\alpha=0.02,0.9$ and $\gamma=0.5,10^{-5}$, as a function of $\beta$.}
    \label{fig:GD2}
\end{figure*} 

The full theory and implementation details of the Graph Decomposition (GD) method are in \cite{stenzel_general_2014}. Here a summary of the GD method and its parameters is provided.
The GD algorithm begins by taking a vertex of minimal vertex degree in $\mathbf{G}$ and uses this vertex as the basis of a cluster. Then each node $v^\prime$ adjacent to $v$ is checked, and added to $\bar{\mathbf{G}}$ if certain criteria are satisfied.
The process is repeated until no more nodes can be added to $\bar{\mathbf{G}}$. At that stage, the nodes of subgraph $\bar{\mathbf{G}}$ are removed from $\mathbf{G}$ and classified as a cluster. The algorithm begins again until all nodes are classified. 

The criteria for including a node into a cluster are:
\begin{enumerate}
    \item 
    Either a completeness criterion or a fullness criterion. 1) The completeness criterion requires that $\frac{R(\bar{\mathbf{G}} \cup \{ v' \} )}{ R(\bar{\mathbf{G}}) } > \alpha$ for some $\alpha > 0$, where $R(\bar{\mathbf{G}})$ is the ratio of the number of edges in the graph $\bar{\mathbf{G}}$ to the number of edges that $\bar{\mathbf{G}}$ would have if it were complete. In the case of $\bar{\mathbf{G}}$ consist of only one node, $R(\bar{\mathbf{G}})=1$ since $\bar{\mathbf{G}}$ is complete. 2) The fullness criterion requires that the external node $v'$ be adjacent to at least a proportion $\beta$ of nodes in the cluster $\bar{\mathbf{G}}$ for some $\beta > 0$.
    \item  A threshold criterion. This requires that at least one transition probability from $v'$ into a node in $\bar{\mathbf{G}}$ be bigger than $\gamma$ and that at least one transition probability from $\bar{\mathbf{G}}$ to $v'$ be bigger than $\gamma$ for some $\gamma > 0$.
\end{enumerate}

Here $\alpha$ characterizes the change in
the ratio of the number of edges in the cluster to the number of edges that G would have if it were complete. 
In our BCP graph, this change is negligible. 
And $\beta$ characterizes the number of edges between a subgraph and an adjacent node. 
In appendix, results of using different $\alpha,\beta$ combinations are tested to identify traps in the BCP structure. Small and large $\gamma$ are used with varying $\alpha,\beta$.
Different $\alpha,\beta$ combinations are tested to identify the single trap node in the BCP structure. Small and large $\gamma$ are used with varying $\alpha,\beta$.
Figure \ref{fig:GD2} shows that at both small $\gamma=10^{-5}$ and large $\gamma=0.5$, various values of $\alpha,\beta$ combination can not identify node $v_1$ as a trap.

\section{Spectral Clustering for non-energy based traps }
\begin{figure}[tbp]
  \centering
  \includegraphics[width=\linewidth]{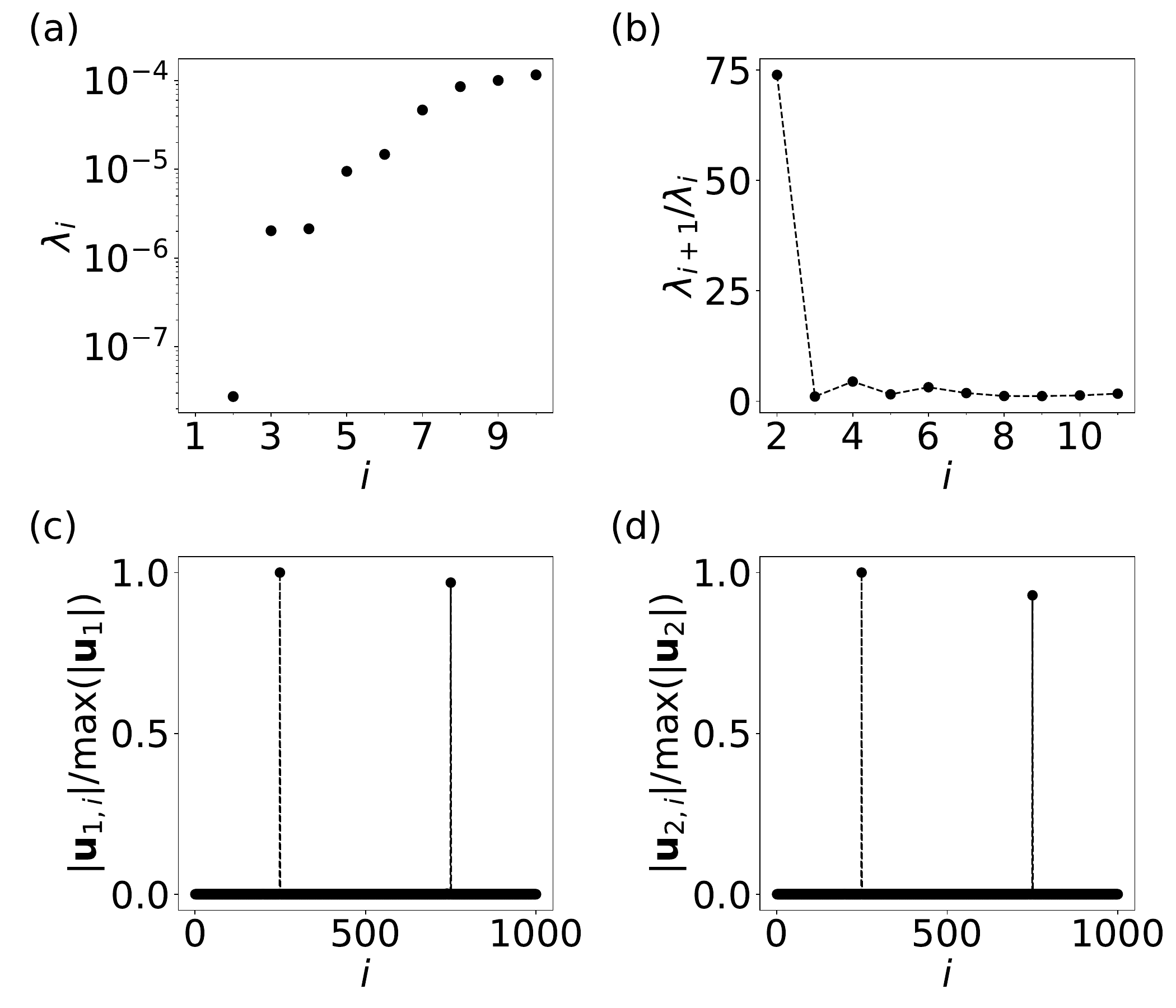}
  \caption{Trap identification results using spectral clustering methods on the BCP system where the site $v_\text{trap}$ and its neighbors have nonsymmetric reorganization energy. (a) The first ten eigenvalues of $L_\text{rw}$. (b) $\lambda_{i+1}/{\lambda_i}$ as a function of $i$. (c)-(d) The first and second normalized eigenvector elements of $L_\text{rw}$ as a function of node indices.}
  \label{fig:lambdaVary}
\end{figure}

\begin{figure}[tbp]
  \centering
  \includegraphics[width=\linewidth]{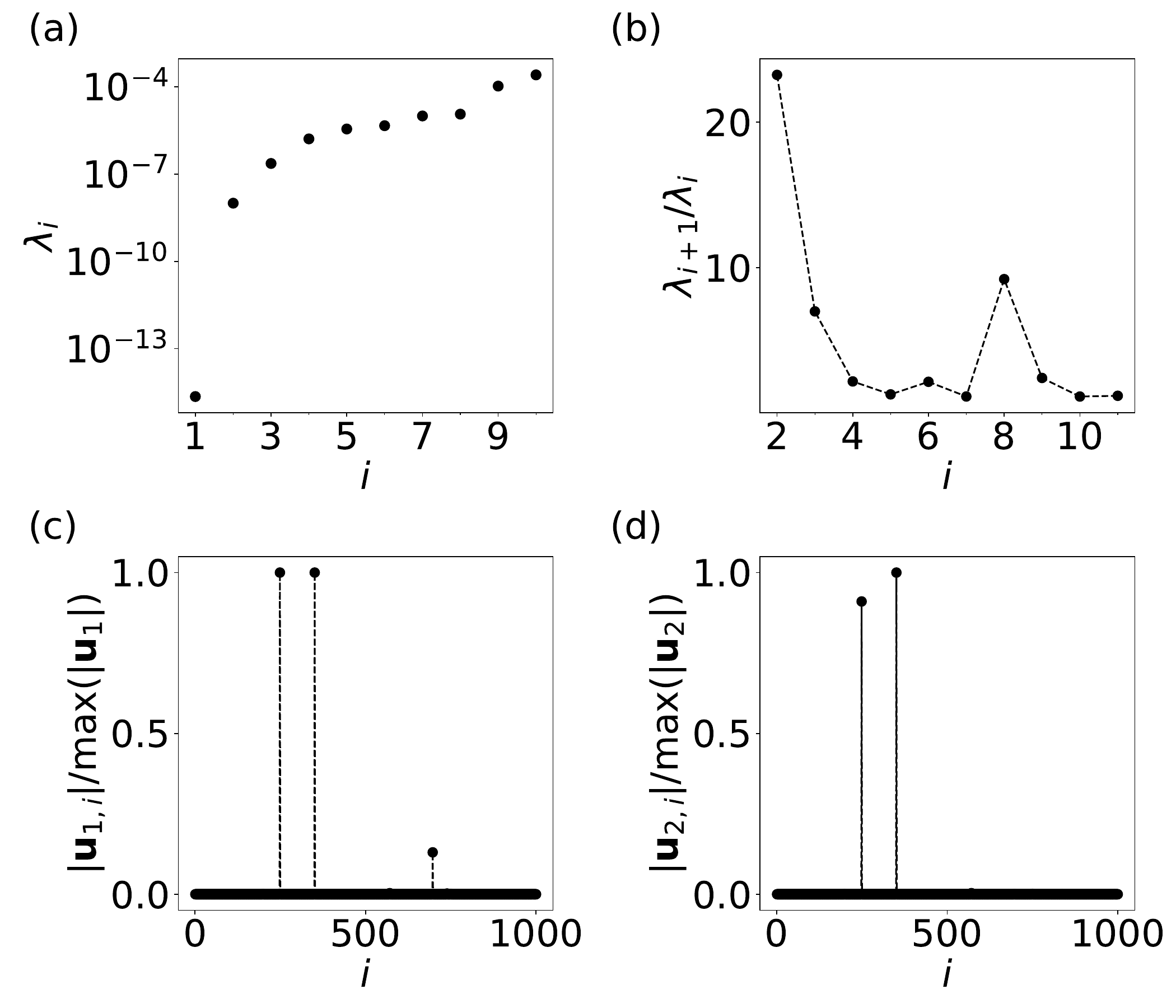}
  \caption{Trap identification results using spectral clustering methods on the BCP system where the node $v_\text{trap}$ and its neighbors have weak coupling elements. (a) The first ten eigenvalues of $L_\text{rw}$. (b) $\lambda_{i+1}/{\lambda_i}$ as a function of $i$. (c)-(d) The first and second normalized eigenvector elements of $L_\text{rw}$ as a function of node indices.}
  \label{fig:Jshield}
\end{figure}

\section*{CRediT author statement}
{\bf Zhongquan Chen}: Conceptualization, Data curation, Formal analysis, Investigation, Methodology, Validation, Visualization, Writing -- original draft, Writing -- review \& editing, {\bf Pim van der Hoorn}: Conceptualization, Methodology, Supervision, Validation, Writing -- original draft, Writing -- review \& editing, {\bf Bj\"orn Baumeier}: Conceptualization, Funding acquisition, Methodology, Resources, Supervision, Validation, Writing -- original draft, Writing -- review \& editing

%

\end{document}